\documentclass[%
reprint, 
amsmath,amssymb,
aps,
pra,
 floatfix,
 nofootinbib
]{revtex4-1}

\usepackage{graphicx,caption}
\usepackage{dcolumn}
\usepackage{bm}
\usepackage{physics}
\usepackage[usenames, dvipsnames]{color}
\usepackage{upgreek}
\usepackage{placeins}
\usepackage{gensymb}
\usepackage{comment}
\usepackage{wasysym}
\usepackage{siunitx}
\usepackage{bbold}
\usepackage{braket}

\usepackage[outline]{contour}

\makeatletter
\newcommand*{\rom}[1]{\expandafter\@slowromancap\romannumeral #1@}
\makeatother

\usepackage[utf8]{inputenc}
\usepackage{bibunits}
\usepackage{natbib}

\AtBeginDocument{\let\oldcontentsline\contentsline}
\newcommand{\notoccontentsline}[4]{\oldcontentsline{#1}{}{}{#4}}
\newcommand{\droptocpage}{\addtocontents{toc}{\let\protect\contentsline\protect\notoccontentsline}}
\newcommand{\incltocpage}{\addtocontents{toc}{\let\protect\contentsline\protect\oldcontentsline}}

\usepackage[skip=2pt,font=footnotesize]{caption}

\begin{document}





\title{Entanglement of dark electron-nuclear spin defects in diamond}

\author{M. J. Degen$^{1,2}$}\thanks{These authors contributed equally to this work.}
\author{S.J.H. Loenen$^{1,2}$}\thanks{These authors contributed equally to this work.}
\author{H. P. Bartling$^{1,2}$}
\author{C. E. Bradley$^{1,2}$}
\author{A.L. Meinsma $^{1,2}$}
\author{M. Markham$^3$} 
\author{D. J. Twitchen$^3$}
\author{T. H. Taminiau$^{1,2}$}
 \email{T.H.Taminiau@TUDelft.nl}

\affiliation{$^{1}$QuTech, Delft University of Technology, PO Box 5046, 2600 GA Delft, The Netherlands}
\affiliation{$^{2}$Kavli Institute of Nanoscience Delft, Delft University of Technology, PO Box 5046, 2600 GA Delft, The Netherlands}
\affiliation{$^{3}$Element Six Innovation, Fermi Avenue, Harwell Oxford, Didcot, Oxfordshire OX11 0QR, United Kingdom}

\date{\today}

\begin{abstract} 
\noindent A promising approach for multi-qubit quantum registers is to use optically addressable spins to control multiple dark electron-spin defects in the environment. While recent experiments have observed signatures of coherent interactions with such dark spins, it is an open challenge to realize the individual control required for quantum information processing. Here we demonstrate the initialisation, control and entanglement of individual dark spins associated to multiple P1 centers, which are part of a spin bath surrounding a nitrogen-vacancy center in diamond. We realize projective measurements to prepare the multiple degrees of freedom of P1 centers - their Jahn-Teller axis, nuclear spin and charge state - and exploit these to selectively access multiple P1s in the bath. We develop control and single-shot readout of the nuclear and electron spin, and use this to demonstrate an entangled state of two P1 centers. These results provide a proof-of-principle towards using dark electron-nuclear spin defects as qubits for quantum sensing, computation and networks.
\end{abstract}

\maketitle


Optically active defects in solids provide promising qubits for quantum sensing \cite{Degen2017QuantumSensing}, quantum-information processing \cite{Waldherr2014QuantumRegister,Cramer2016RepeatedFeedback,Bradley2019AMinute}, quantum simulations \cite{Cai2013ATemperature,Wang2015QuantumRegister} and quantum networks \cite{Bernien2013HeraldedMetres, HensenLoophole-freeKilometres,Sipahigil2016AnNetworks}. These defects, including the nitrogen-vacancy (NV) and silicon-vacancy (SiV) centers in diamond and various defects in silicon-carbide \cite{Awschalom2018QuantumSpins,Atature2018MaterialTechnologies, Castelletto2020SiliconApplications}, combine long spin coherence times \cite{Abobeih2018One-secondEnvironment,Bradley2019AMinute,Sukachev2017Silicon-VacancyReadout, Rose2018ObservationFrom,Nagy2018QuantumCarbide,Simin2017LockingCarbide,bar2013solid}, high-quality control and readout \cite{Cramer2016RepeatedFeedback, Bradley2019AMinute,Waldherr2014QuantumRegister, Sukachev2017Silicon-VacancyReadout, Nagy2019High-fidelityCarbide, Koehl2011RoomCarbide, Maity2020CoherentDiamond}, and a coherent optical interface \cite{Bernien2013HeraldedMetres,HensenLoophole-freeKilometres, Sipahigil2016AnNetworks,Rose2018ObservationFrom, Green2019ElectronicDiamond, Nagy2019High-fidelityCarbide}. 

Larger-scale systems can be realized by entangling multiple defects together through long-range optical network links \cite{Sipahigil2016AnNetworks,Bernien2013HeraldedMetres, HensenLoophole-freeKilometres} and through direct magnetic coupling, as demonstrated for a pair of ion-implanted NV centers \cite{Dolde2013Room-temperatureDiamond, Dolde2014High-fidelityControl}. The number of available spins can be further extended by controlling nuclear spins in the vicinity. Multi-qubit quantum registers \cite{Dolde2014High-fidelityControl,Bradley2019AMinute, VanDam2019MultipartiteMeasurements,Unden2019RevealingCenters,Hou2019ExperimentalRegister}, quantum error correction \cite{Cramer2016RepeatedFeedback,Waldherr2014QuantumRegister}, enhanced sensing schemes \cite{Vorobyov2020QuantumSensing}, and entanglement distillation \cite{Kalb2017EntanglementNodes} have been realized using nuclear spins.  

The ability to additionally control dark electron-spin defects that cannot be directly detected optically would open new opportunities. Examples are studying single defect dynamics \cite{Yang2016ProposalImprovin}, extended quantum registers, enhanced sensing protocols \cite{Cooper2019Environment-assistedDiamond, Goldstein2011Environment-AssistedMeasurement, Vorobyov2020QuantumSensing} and spin chains for quantum computation architectures \cite{Yao2012ScalableProcessor, Ping2013PracticalityTechnologies, Yao2013QuantumRegisters,Schlipf2017AQubit}. Two pioneering experiments reported signals consistent with an NV center coupled to a single P1 center (a dark substitutional nitrogen defect) \cite{Gaebel2006Room-temperatureDiamond,Hanson2006PolarizationDiamond}, but the absence of the expected P1 electron-spin resonance signal \cite{HansonWrachtrupPrivComm} and later results revealing identical signals due to NV-$^{13}$C couplings in combination with an excited state anti-crossing \cite{dreau2012high}, make these assignments inconclusive. Recent experiments have revealed signatures of coherent interactions between NV centers and individual dark electron-spin defects, including P1 centers \cite{Knowles2014, Knowles2016b, Belthangady2013Dressed-stateDiamond}, N2 centers \cite{Shi2013QuantumSpin} and not-yet-assigned defects \cite{Cooper2019Environment-assistedDiamond,Cooper2020, Yamamoto2013StronglyImplantation, Rosenfeld2018, Grinolds2014SubnanometreSpins,sun2020improved}. Those results have revealed the prospect of using dark spin defects as qubits. However, high-quality initialisation, measurement and control of multi-qubit quantum states is required to exploit such spins as a quantum resource.

\begin{figure*}[tb]
\includegraphics[scale=0.9]{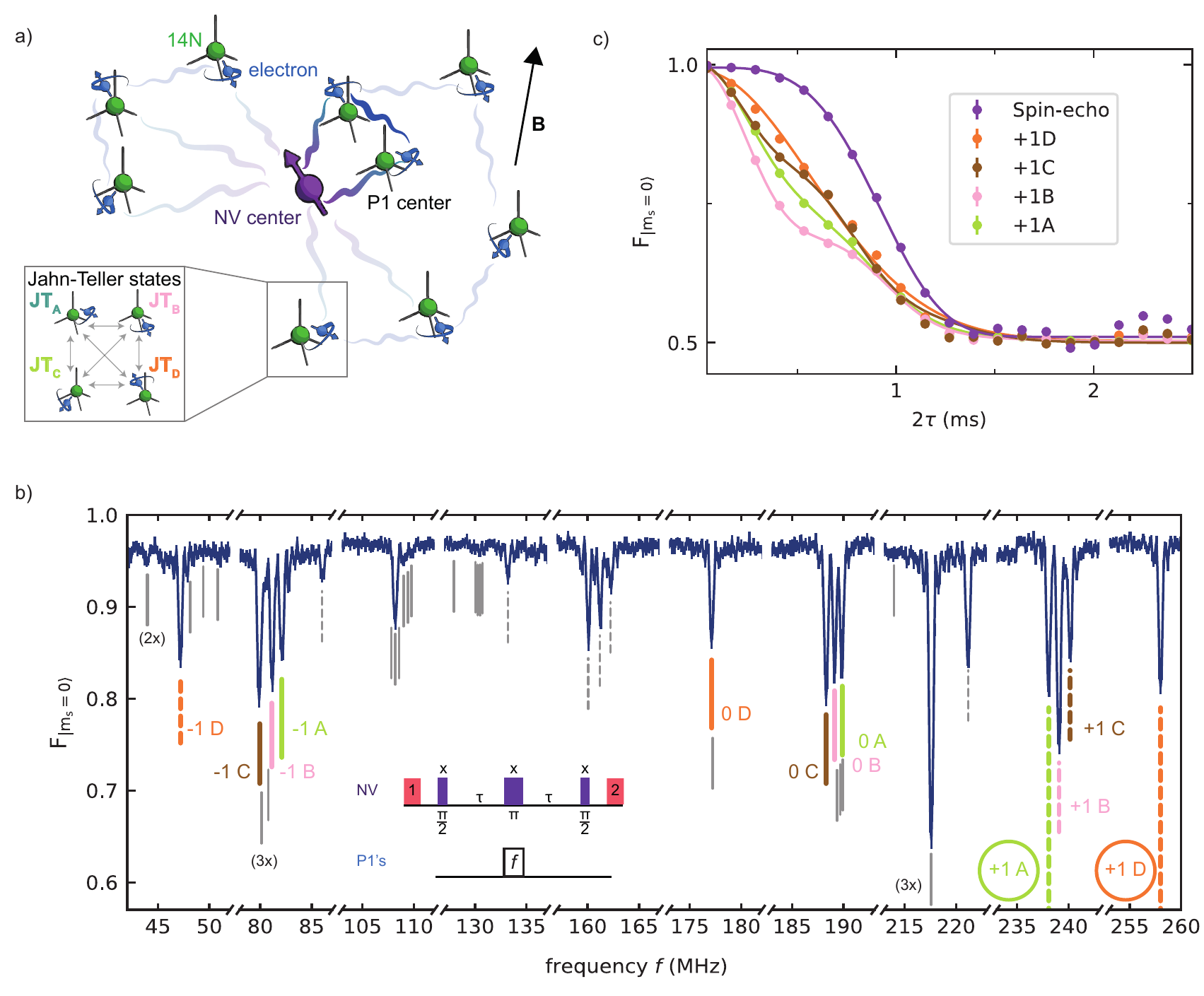}
\caption{\label{Figure1}
\textbf{DEER spectroscopy of a P1 spin bath.} \textbf{a)} We study a bath of P1 centers surrounding a single NV center. The  state  of each P1 center is defined by an electron spin (blue), a $^{14}$N nuclear spin (green), and one of four JT axis, which can vary over time (see inset). \textbf{b)} DEER spectrum obtained by varying the frequency $f$ (see inset). The NV is initialized in $m_s$ = 0 via spin-pumping (1) and read out (2) at the end of the sequence (Methods). $F_{\ket{m_s=0}}$ is the fidelity of the final NV state with $m_s=0$. The 12 main P1 electron-spin transitions are labelled by their nitrogen state and JT axis (colored lines). 11 isolated transitions (dashed lines) are used to fit the P1 Hamiltonian and all predicted transition frequencies are indicated (solid lines). In this work, we mainly use the circled transitions corresponding to $\ket{+1,\mathrm{D}}$ and $\ket{+1,\mathrm{A}}$. \textbf{c)} We apply a calibrated $\pi$ pulse (Rabi frequency $\Omega$ = 250 kHz) at a fixed frequency $f$, to selectively couple to P1 centers in the $\ket{+1, i}$ state ($i$ $\in$ $\{$A,B,C,D$\}$) and vary the interaction time $2\tau$ (see inset in b). From the fits we obtain a dephasing time $T_{2,DEER}$ of 0.767(6), 0.756(7), 0.802(6) and 0.803(5) ms for the $\ket{+1, i}$ state with $i$ corresponding to A,B,C and D respectively.  A spin-echo (no pulse on P1 centers) is added for reference from which we obtain $T_{2,NV}$ = 0.992(4) ms. Error bars are one standard deviation (Methods), with a typical value $\num{4e-3}$, which is smaller than the data points. See Methods for the fit functions.}
\end{figure*}

Here we demonstrate the control and entanglement of individual P1 centers that are part of a bath surrounding an NV center in diamond (Fig. \ref{Figure1}a). A key property of the P1 center is that, in addition to its electron spin, it exhibits three extra degrees of freedom: the Jahn-Teller axis, a nuclear spin, and the charge state \cite{smith1959electron, Ulbricht2011SingleSpectroscopy, Deak2014FormationDefects}. Underlying our selective control of individual centers is the heralded preparation of specific configurations of these additional degrees of freedom for multiple P1 centers through projective measurements. In contrast, all previous experiments averaged over these additional degrees of freedom \cite{DeLange2012,Knowles2014,Knowles2016b}. We use this capability to develop initialisation, single-shot readout and control of the electron and nuclear spin states of multiple P1s, and investigate their spin relaxation and coherence times. Finally, we demonstrate the potential of these dark spins as a qubit platform by realizing an entangled state between two P1 electron spins through their direct magnetic-dipole coupling.\\

\noindent\textbf{RESULTS}

{\bf{A spin bath with multiple degrees of freedom.}} 
 
\noindent We consider a bath of P1 centers surrounding a single NV center at 3.3 K (Fig. 1a). The diamond is isotopically purified with an estimated $^{13}$C concentration of $0.01 \%$. The P1 concentration is estimated to be $\sim$ 75 ppb (see Supplementary Note \rom{5}). Three P1 charge states are known \cite{Ulbricht2011SingleSpectroscopy,Deak2014FormationDefects}. The experiments in this work detect the neutral charge state and do not generate signal for the positive and negative charge states. In addition to an electron spin ($S=1/2$), the P1 center exhibits a $^{14}$N nuclear spin ($I=1$, $99.6\%$ natural abundance) and a Jahn-Teller (JT) distortion, which results in four possible symmetry axes due to the elongation of one of the four N-C bonds \cite{Loubser1978}. Both the $^{14}$N state and the JT axis generally fluctuate over time \cite{Konchits1977,Of1967, Ammerlaan1981}. The Hamiltonian for a single neutrally-charged P1 defect in one of the four JT axis $i$ $\in$ $\{$A,B,C,D$\}$ is \cite{smith1959electron}:

\begin{equation}
    H_{\bf i, \text{P1}} = \gamma_e \vec{B}\cdot\vec{S} + \gamma_n \vec{B}\cdot\vec{I} + \vec{I}\cdot {\bf \hat{P}_i}\cdot \vec{I} + \vec{S}\cdot {\bf \hat{A}_i}\cdot \vec{I},
    \label{eq:1}
\end{equation}

\begin{figure*}[tb]
\includegraphics[scale=0.9]{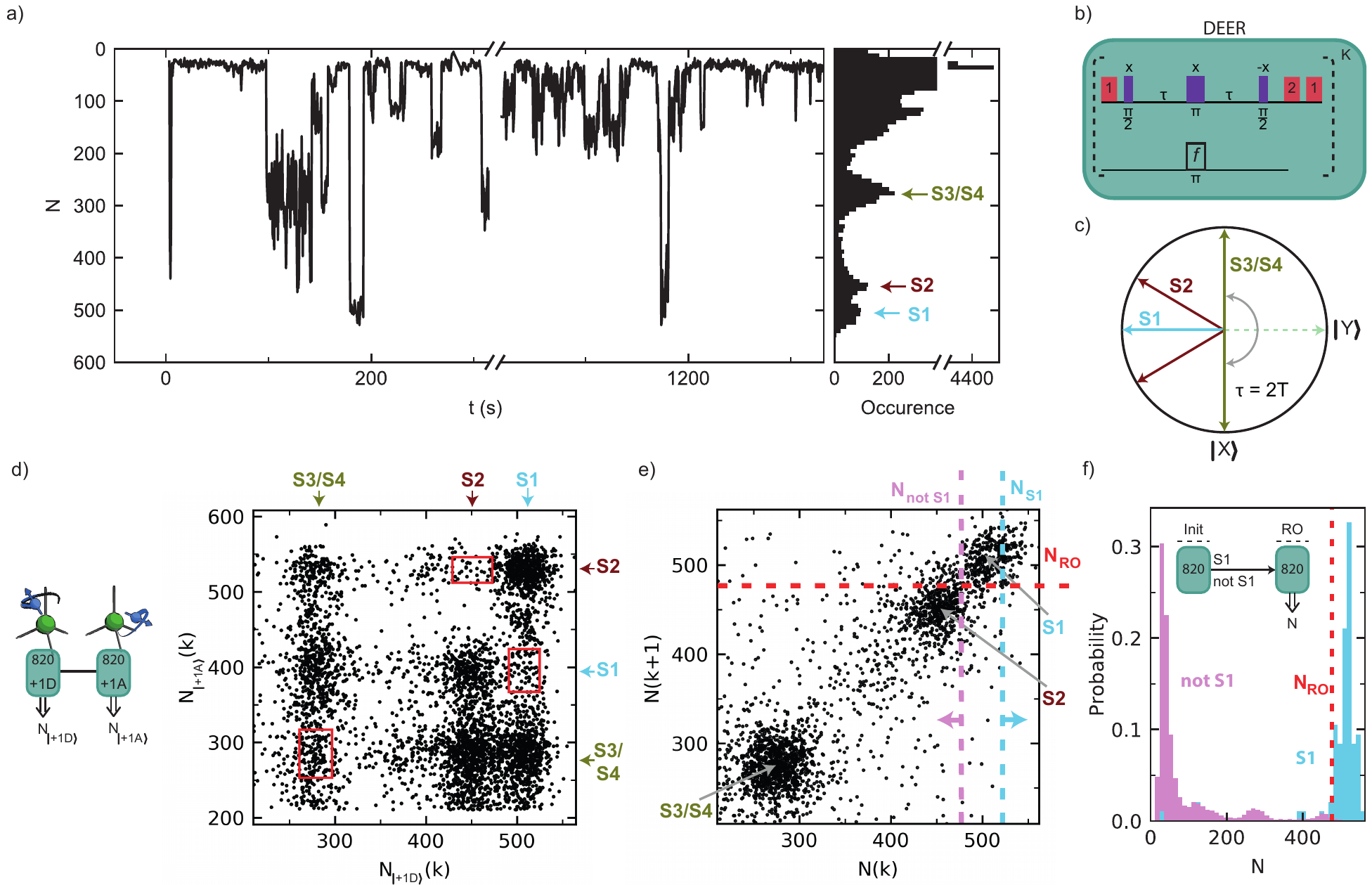}
\caption{\label{Figure2} \textbf{Detection and preparation of single P1 centers.} \textbf{a)} Typical time trace for the DEER signal for $\ket{+1,\mathrm{D}}$. $N$ is the total number of $m_s=0$ NV readout outcomes in $K=820$ repetitions of the sequence (see (b)). The discrete jumps and corresponding peaks in the histogram of the full time trace ($\sim$ 6 h, right) indicate that several individual P1s are observed (S1,  S2 and S3/S4). \textbf{b)} Sequence for $K$ repeated DEER measurements. \textbf{c)} XY-plane of the NV-spin Bloch sphere before the second $\pi$/2 pulse of a DEER measurement, with the NV initialised along +z at the start. The NV spin picks up phase depending on which nearby P1 centers are in the targeted $\ket{+1,\mathrm{D}}$ state. Because the NV spin is effectively measured along the y-axis, this sequence is insensitive to the P1 electron spin state. \textbf{d)} Cross-correlation of two consecutive DEER measurements for $\ket{+1,\mathrm{D}}$ ($K$=820) and $\ket{+1,\mathrm{A}}$ ($K$=820). Three areas (red boxes, Supplementary Note \rom{8}) show an anti-correlation associated to S1, S2, and S3/S4, in agreement with the assignment of discrete P1 centers. Left: sequence for the two consecutive DEER measurements (green blocks). Double lined arrows indicate measurement outcomes. \textbf{e)} Correlation plot for consecutive measurement outcomes $N(k)$ and $N(k+1)$, both for $\ket{+1,\mathrm{D}}$. Dashed lines are the thresholds used to prepare (vertical) and read out (horizontal) the JT and $^{14}$N state in panel f. We use $N_{S1}$ $>$ 522 to prepare S1 in $\ket{+1,\mathrm{D}}$, and S2 and S3/S4 in any other state. The condition $N_{notS1}$ $\leq$ 477 prepares a mixture of all other possibilities. A threshold $N_{RO}$ = 477 distinguishes between those two cases in readout. \textbf{f)} Conditional probability distributions for both preparations, demonstrating initialisation and single-shot readout of the $^{14}$N and JT state of S1. Inset: experimental sequence. Labelled horizontal arrows indicate conditions for passing the initialisation measurement (init).}
\end{figure*}

\noindent where $\gamma_e$ ($\gamma_n$) is the electron ($^{14}$N) gyromagnetic ratio, $\vec{B}$ the external magnetic field vector, $\vec{S}$ and $\vec{I}$ are the electron spin-1/2 and nuclear spin-1 operator vectors, and ${\bf \hat{A}_i}$ (${\bf \hat{P}_i}$) the hyperfine (quadrupole) tensor. We label the $^{14}$N ($m_I \in {-1,0,+1}$) and JT states as $\ket{m_I,i}$, and the electron spin states as $\ket{\uparrow}$ and $\ket{\downarrow}$. For convenience, we use the spin eigenstates as labels, while the actual eigenstates are, to some extent, mixtures of the $^{14}$N and electron spin states.

We probe the bath surrounding the NV by double electron-electron resonance (DEER) spectroscopy \cite{Knowles2016b,Knowles2014,Rosenfeld2018,Cooper2020,DeLange2012}. The DEER sequence consists of a spin echo on the NV electron spin, which decouples it from the environment, plus a simultaneous $\pi$-pulse that selectively recouples resonant P1 centers. Figure \ref{Figure1}b reveals a complex spectrum. The degeneracy of three of the JT axes is lifted by a purposely slightly tilted magnetic field with respect to the NV axis ($\theta$ $\approx$ 4$^{\circ}$). In combination with the long P1 dephasing time ($T_2^* \sim\ 50~\mathrm{\upmu}$s, see Fig \ref{Figure4}d) this enables us to resolve all 12 main P1 electron-spin transitions -- for four JT axes and three $^{14}$N states -- and selectively address at least one transition for each JT axis.

Several additional transitions are visible due to mixing of the electron and nuclear spin in the used magnetic field regime ($\gamma_e|\vec{B}|\sim A_{\parallel}$,$A_{\bot}$). We select 11 well-isolated transitions to fit the P1 Hamiltonian parameters and obtain $\{$\textit{A{$_{\parallel}$}}, \textit{A{$_{\bot}$}}, \textit{P$_{\parallel}$}$\}$ = $\{$114.0264(9), 81.312(1), -3.9770(9)$\}$ MHz and $\vec{B}$ = $\{$2.437(2), 1.703(1), 45.5553(5)$\}$ G (Supplementary Note \rom{4}), closely matching ensemble ESR measurements \cite{cook1966electron}. The experimental spectrum is well described by the 60 P1 transitions for these parameters. No signal is observed at the bare electron Larmor frequency ($\approx$ 128 MHz), confirming that the P1 centers form the dominant electron spin bath.     

To probe the coupling strength of the P1 bath to the NV, we sweep the interaction time in the DEER sequences (Fig. \ref{Figure1}c). The curves for the different $\ket{+1,i}$ states show oscillatory features, providing a first indication of an underlying microscopic structure of the P1 bath. However, like all previous experiments \cite{DeLange2012,Knowles2014,Knowles2016b}, these measurements are a complex averaging over $^{14}$N, JT and charge states for all the P1 centers, which obscures the underlying structure and hinders control over individual spins. 

{\bf{Detecting and preparing single P1 centers.}}

\noindent To investigate the microscopic structure of the bath we repeatedly apply the DEER sequence and analyze the correlations in the measurement outcomes \cite{Yang2016ProposalImprovin}. Figure $\ref{Figure2}$a shows a typical time trace for continuous measurement, in which groups of $K$=820 measurements are binned together (see Fig. $\ref{Figure2}$b for the sequence). We observe discrete jumps in the signal that indicate individual P1 centers jumping in and out of the $\ket{+1,\mathrm{D}}$ state. The resulting histogram (Fig. $\ref{Figure2}$a) reveals multiple discrete peaks that indicate several P1 centers with different coupling strengths to the NV center, as schematically illustrated in Fig. $\ref{Figure2}$c. We tentatively assign four P1 centers S1, S2, S3 and S4 to these peaks.

\begin{figure}
\includegraphics[scale=0.9]{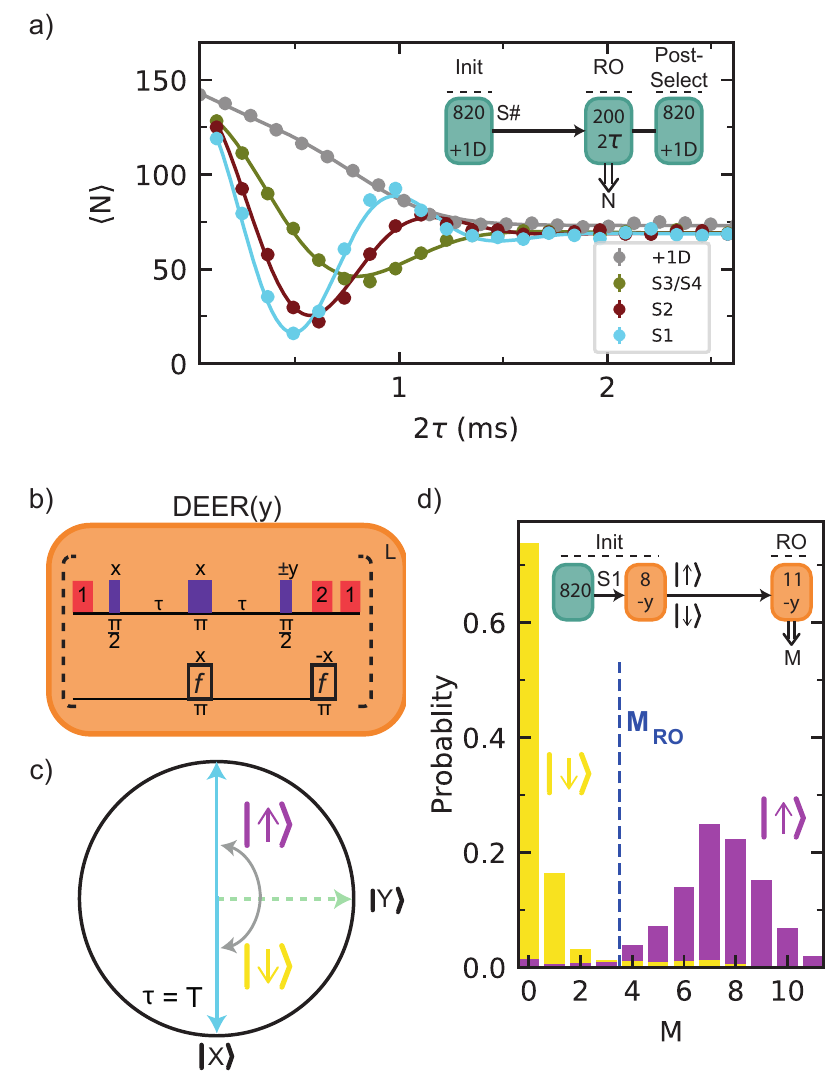}
\caption{\label{Figure3}
\textbf{Electron spin initialisation and readout.} \textbf{a)} Measuring the NV-P1 coupling strength. We initialize S1, S2, or S3/S4 in $\ket{+1,\mathrm{D}}$ and vary the interaction time 2$\tau$ of a DEER sequence. $\langle N\rangle$ is the mean of the number of NV $m_s=0$ outcomes for $K$=200 repetitions. To improve the signal, the results are post-selected on again obtaining $\ket{+1,\mathrm{D}}$. Error bars are one standard deviation (Methods), with a typical value $1$, which is smaller than the data points. Grey: without P1 initialisation (data from Fig. \ref{Figure1}c). \textbf{b)} DEER(y) sequence with the readout basis rotated by $\pi/2$ compared to the DEER sequence and $\tau$ = $\pi$/$2\nu$. An additional $\pi$ pulse is added to revert the P1 electron spin. \textbf{c)} XY-plane of the NV Bloch sphere before the second $\pi$/2 pulse, illustrating that the DEER(y) sequence measures the P1 electron spin state (shown for positive NV-P1 coupling). \textbf{d)}  Single-shot readout of the S1 electron spin. After preparation in $\ket{+1,\mathrm{D}}$, the electron spin is initialized through a DEER(y) measurement ($L$=8) with thresholds $M_{\ket{\uparrow}}$ ($>6$) and $M_{\ket{\downarrow}}$ ($\leq$1). Shown are the conditional probability distributions for a subsequent DEER(y) measurement with $L$=11 and the readout threshold $M_{RO}$.} 
\end{figure}

We verify whether these peaks originate from single P1 centers by performing cross-correlation measurements. We first apply a DEER measurement on $\ket{+1,\mathrm{D}}$ followed by a measurement on $\ket{+1,\mathrm{A}}$ (Fig. $\ref{Figure2}$d). For a single P1, observing it in $\ket{+1,\mathrm{D}}$ would make it unlikely to subsequently also find it in state $\ket{+1,\mathrm{A}}$. We observe three regions of such anti-correlation (red rectangles in Fig. \ref{Figure2}d). We define the correlation:

\begin{equation}
\scalebox{1.1}{
    $ C = \frac{P( N_A^{min} \leq N_{\ket{+1,\mathrm{A}}} \leq N_{A}^{max}  \big|   N_{D}^{min} \leq N_{\ket{+1,\mathrm{D}}} \leq N_{D}^{max})}{P(N_{A}^{min} \leq N_{\ket{+1,\mathrm{A}}} \leq N_{A}^{max})},$}
\end{equation}
where $N_A^{min}$, $N_A^{max}$, $N_D^{min}$ and $N_D^{max}$ define the region, and where $P(X)$ is the probability that $X$ is satisfied. Assuming that the states of different P1 centers are uncorrelated, a value $C$ $<$ 0.5 indicates that the signal observed in both the DEER sequences on $\ket{+1,\mathrm{A}}$ and  $\ket{+1,\mathrm{D}}$ is associated to a single P1 center, while $C<2/3$ indicates 1 or 2 centers (Supplementary Note \rom{8}). 

For the three areas we find $C=0.40(5)$, $0.22(4)$ and $0.47(5)$ for S1, S2 and S3/S4 respectively. These correlations corroborate the assignments of a single P1 to both S1 and S2 and one or two P1s for S3/S4 (the result is within one standard deviation from 0.5). Additionally, these results reveal which signals for different $\ket{+1,i}$ states belong to which P1 centers. This is non-trivial because the NV-P1 dipolar coupling varies with the JT axis, as exemplified in Fig. $\ref{Figure2}$d (see Supplementary Note \rom{3} for a theoretical treatment).


Next, we develop single-shot readout and initialisation of the $^{14}$N and JT state of individual P1 centers. For this, we represent the time trace data (Fig. \ref{Figure2}a) as a correlation plot between subsequent measurements $k$ and $k+1$ (Fig. $\ref{Figure2}$e) \cite{dreau2013single, liu2017single, neumann2010single}. We bin the outcomes using $K$=820 repetitions as a trade-off between the ability to distinguish S1 from S2 and the disturbance of the state due to the repeated measurements (1/e value of $\sim$\num{1.5e4} repetitions, see Supplementary Note \rom{6}). Separated regions are observed for the different P1 centers. Therefore, by setting threshold conditions, one can use the DEER measurement as a projective measurement to initialize or readout the $\ket{m_I,i}$ state of selected P1 centers, which we illustrate for S1. 

First, we set an initialisation condition $N(k) > N_{S1}$ (blue dashed line) to herald that S1 is initialized in the $\ket{ +1, \mathrm{D}}$ state and that S2, S3/S4 are not in that state. We use $N(k) \leq N_{not S1}$ to prepare a mixture of all other other possibilities. The resulting conditional probability distributions of $N(k+1)$ are shown in Fig. 2f. Second, we set a threshold for state readout $N_{RO}$ to distinguish between the two cases. We then optimize $N_{S1}$ for the trade-off between the success rate and signal contrast, and find a combined initialisation and readout fidelity $F$ = 0.96(1) (see Methods). Other states can be prepared and read out by setting different conditions (Supplementary Note \rom{8}).

{\bf{Control of the electron and nuclear spin.}}

To control the electron spin of individual P1 centers, we first determine the effective dipolar NV-P1 coupling. We prepare, for instance, S1 in $\ket{+1, \mathrm{D}}$ and perform a DEER measurement in which we sweep the interaction time (Fig. \ref{Figure3}a). By doing so, we selectively couple the NV to S1, while decoupling it from S2 and S3/S4, as well as from all bath spins that are not in $\ket{+1, \mathrm{D}}$. By applying this method we find effective dipolar coupling constants $\nu$ of $2\pi \cdot$1.910(5), $2\pi \cdot$1.563(6) and $2\pi \cdot$1.012(8) kHz for S1, S2 and S3/S4 respectively. Note that, if the signal for S3/S4 originates from two P1 centers, the initialisation sequence prepares either S3 or S4 in each repetition of the experiment.

We initialize and measure the electron spin state of the P1 centers through a sequence with a modified readout axis that we label DEER(y) (Fig. 3b). Unlike the DEER sequence, this sequence is sensitive to the P1 electron spin state. After initializing the charge, nuclear spin and JT axis, and setting the interaction time $\tau \approx \pi/ (2\cdot\nu)$, the DEER(y) sequence projectively measures the spin state of a selected P1 center (Fig. 3c). We first characterize the P1 electron spin relaxation under repeated application of the measurement and find a 1/e value of $\sim$250 repetitions (Supplementary Note \rom{4}). We then optimize the number of repetitions and the initialisation and readout thresholds to obtain a combined initialisation and single-shot readout fidelity for the S1 electron spin of $F_{\ket{\uparrow}/\ket{\downarrow}}$ = 0.95(1) (Fig. 3d).

\begin{figure}
\includegraphics[scale=0.9]{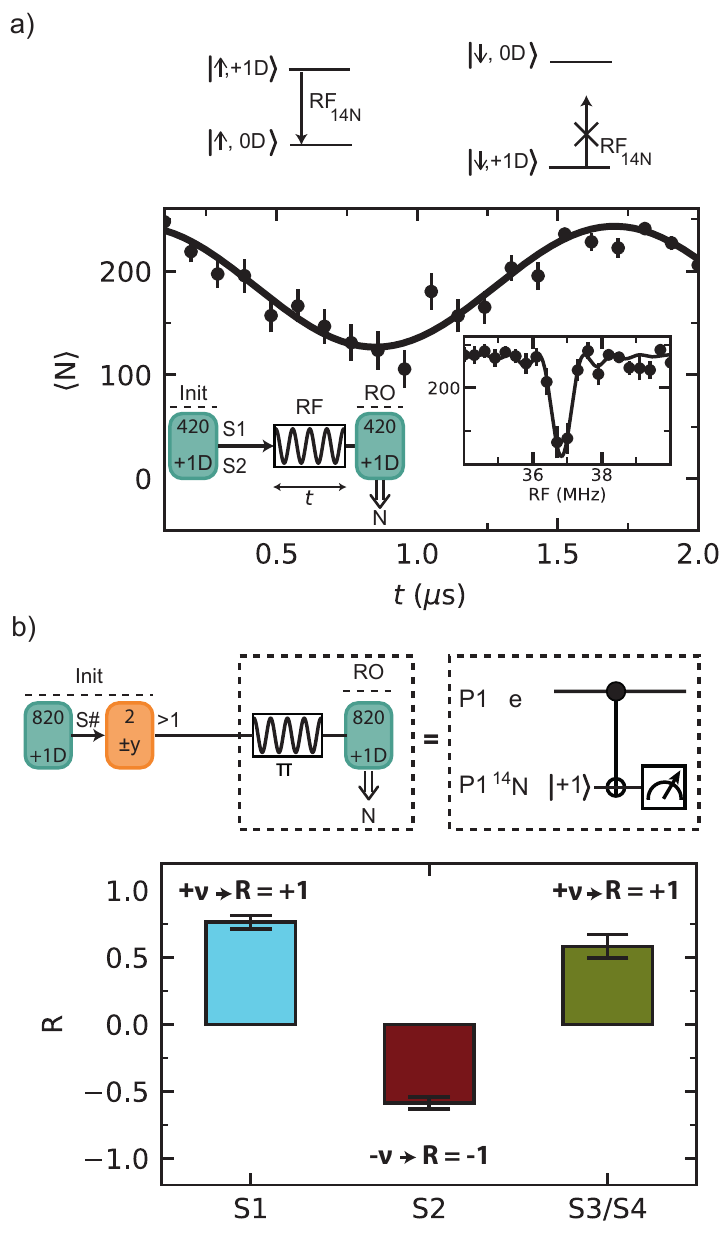}
\caption{\label{Figure3_partII}
\textbf{Nitrogen spin control and NV-P1 coupling sign.} \textbf{a)} $^{14}$N Rabi oscillation. Top: energy levels of the P1 electron spin in the $\{ 0\mathrm{D}, +1\mathrm{D} \}$ subspace. Bottom: either S1 or S2 is prepared in $\ket{+1,\mathrm{D}}$ and the length $t$ of a pulse at frequency $RF$ = $RF_{14N} = 36.8$  MHz is varied. The nitrogen spin is driven conditionally on the electron spin state. Inset: NMR spectrum obtained by varying the frequency $RF$ for a fixed pulse duration $t$. \textbf{b)} We use the $^{14}$N spin to determine the sign of the NV-P1 coupling. First, we prepare S1 or S2 ($K$=820) and initialise their electron spin ($L$=2). Second, we apply a $\pi$ pulse at $RF_{14N}$, which implements an electron controlled CNOT$_{\mathrm{e},\mathrm{N}}$ (see level structure in (a)). The coupling sign to the NV determines the P1 electron-spin state, and, in turn, the final $^{14}$N state. Finally, we measure the fidelity with the $^{14}$N $\ket{+1}$ state for two opposite electron spin initialisations (+y and -y final $\pi$/2 pulse of DEER(y)). The normalized difference $R$ of these measurements reveals the sign of the coupling (see Methods). All error bars indicate one statistical standard deviation.} 
\end{figure}

\begin{figure*}[tb]
\includegraphics{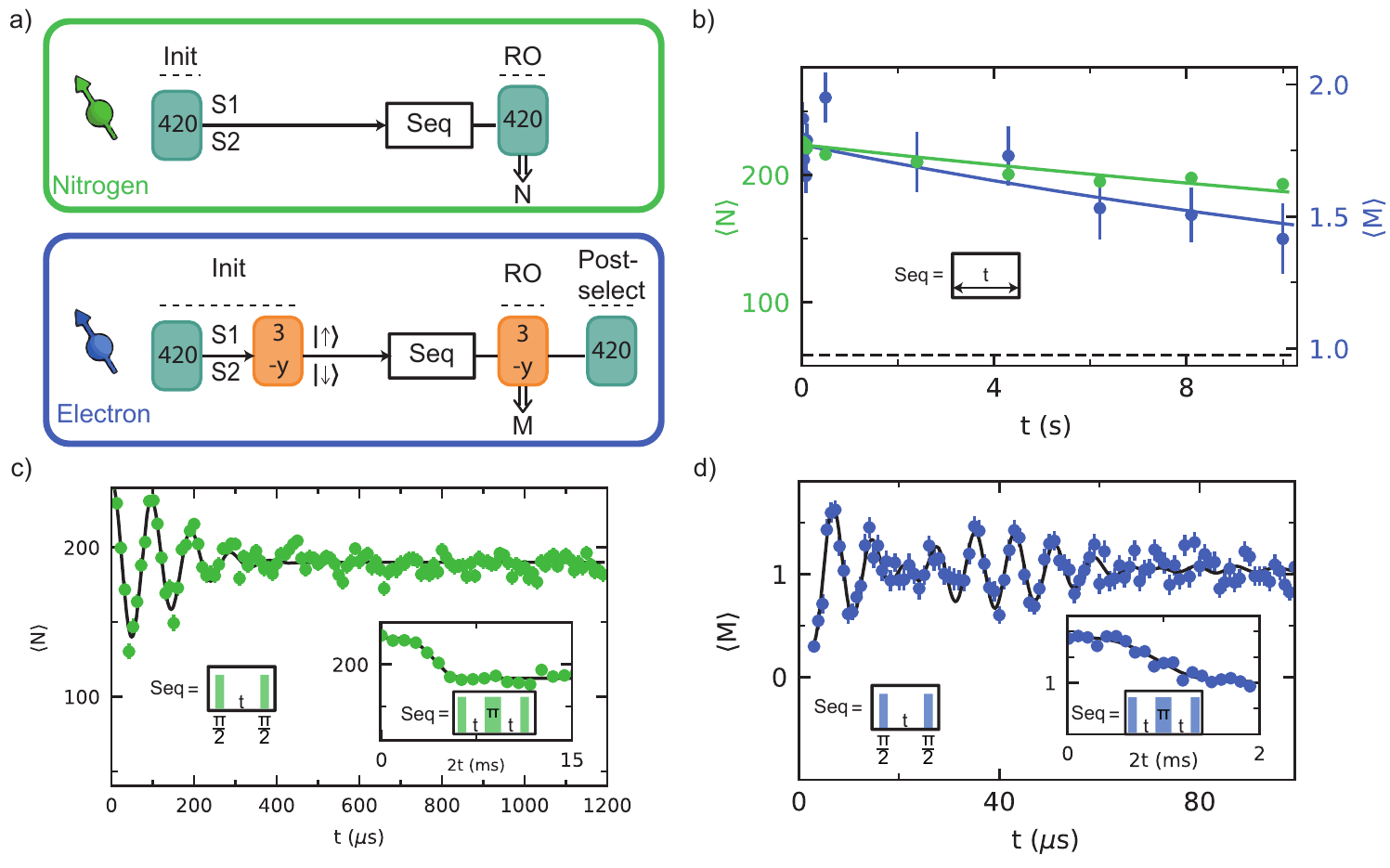}
\caption{\label{Figure4}
\textbf{Coherence and timescales.} \textbf{a)} Sequence for initialisation of either S1 or S2 in $\ket{+1,\mathrm{D}}$ (top). Sequence for initializing all degrees of freedom of either S1 or S2, including the electron spin state (bottom). These sequences are used in b, c and d. \textbf{b)} Relaxation of a combination of: the nitrogen state, JT axis and charge state (green), and only the electron spin state (blue). We fit (solid lines) both curves to $o + A_0 e^{-t/T}$, where $o$ is fixed to the uninitialized mean value (dashed line) and obtain $T$=$T_{\ket{+1,\mathrm{D}}}$ = 40(4) s and $T$=$T_1$ = 21(7) s. \textbf{c)} Ramsey experiment on the nitrogen spin. We fit the data (solid line) and obtain $T^*_{2 N}$ = 0.201(9) ms. (inset) Nitrogen spin-echo experiment. From the fit we obtain $T_{2 N}$ = 4.2(2) ms. \textbf{d)} Ramsey experiment on the electron spin. A Gaussian decay ($T^*_{2 e}$ = 50(3) $\upmu$s) with a single beating is observed, suggesting a dipolar coupling between S1 and S2. (inset) Electron spin-echo experiment. From the fit we obtain $T_{2e}$ = 1.00(4) ms. See Methods for complete fit functions and obtained parameters. All error bars indicate one statistical standard deviation. }

\end{figure*}

We now show that we can coherently control the P1 nitrogen nuclear spin (Fig. \ref{Figure3_partII}a). To speed up the experiment, we choose a shorter initialisation sequence that prepares either S1 or S2 in the $\ket{+1,\mathrm{D}}$ state. We then apply a radio-frequency (RF) pulse that is resonant with the $m_I = $ +1 $\leftrightarrow$ 0 transition if the electron spin is in the $\ket{\uparrow}$ state. Varying the RF pulse length reveals a coherent Rabi oscillation. Because the P1 electron spin is not polarized, the RF pulse is on resonance $50\%$ of the time and the amplitude of the Rabi oscillation is half its maximum.

We use the combined control over the electron and nuclear spin to determine the sign of the NV-P1 couplings (Fig. \ref{Figure3_partII}b). First, we initialize the $^{\mathrm{14}}$N, JT axis and electron spin state of a P1 center. Because the DEER(y) sequence is sensitive to the sign of the coupling (Fig. 3c), the sign affects whether the P1 electron spin is prepared in $\ket{\uparrow}$ or $\ket{\downarrow}$. Second, we measure the P1 electron spin through the $^{14}$N nuclear spin. We apply an RF pulse, which implements an electron-controlled CNOT gate on the nuclear spin (see Fig. \ref{Figure3_partII}a). Subsequently reading out the $^{14}$N spin reveals the electron spin state and therefore the sign of the NV-P1 coupling. We plot the normalized difference $R$ (Methods) for two different initialisation sequences that prepare the electron spin in opposite states. The results show that NV-P1 coupling is positive for the cases of S1 and S3/S4, but negative for S2 (Fig. \ref{Figure3_partII}b). If S3/S4 consists of two P1 centers, then they have the same coupling sign to the NV.

{\bf{Spin coherence and relaxation.}}

To assess the potential of P1 centers as qubits, we measure their coherence times. First, we investigate the relaxation times. We prepare either S1 or S2 in $\ket{+1,\mathrm{D}}$, the NV electron spin in $m_s= 0$, and vary the waiting time $t$ before reading out the same state (Fig. \ref{Figure4}a). This sequence measures the relaxation of a combination of the nitrogen spin state, JT axis and charge state, averaged over S1 and S2. An exponential fit gives a relaxation time of $T_{\ket{+1,\mathrm{D}}}$ = 40(4) s (Fig. \ref{Figure4}b, green). 

We measure the longitudinal relaxation of the electron spin by preparing either $\ket{\uparrow}$ (S1) or $\ket{\downarrow}$ (S2) (Fig. \ref{Figure4}a). We post-select on the $\ket{+1,\mathrm{D}}$ state at the end of the sequence to exclude effects due to relaxation from $\ket{+1,\mathrm{D}}$, and find  $T_{1e}$ = 21(7) s. The observed electron spin relaxation time is longer than expected from the typical P1-P1 couplings in the bath (order of 1 kHz). A potential explanation is that flip-flops are suppressed due to couplings to neighbouring P1 centers, which our heralding protocol preferentially prepares in other $\ket{m_I,i}$ states. Below, we will show that S1 and S2 have a strong mutual coupling, which could shift them off resonance from the rest of the bath. 

Second, we investigate the electron and nitrogen spin coherence via Ramsey and spin-echo experiments (Figs. \ref{Figure4}c and d). We find $T^*_{2e}$ = 50(3) $\upmu$s and $T_{2e}$ = 1.00(4) ms for the electron spin, and $T^*_{2 N}$ = 0.201(9) ms and $T_{2 N}$ = 4.2(2) ms for the nitrogen spin. The ratio of dephasing times for the electron and nitrogen spins is $\sim$4, while the difference in bare gyromagnetic ratios is a factor $\sim$9000. The difference is partially explained by electron-nuclear spin mixing due to the large value of $A_{\bot}$, which changes the effective gyromagnetic ratios of the $^{14}N$ and electron spin. Based on this, a ratio of dephasing times of $12.6$ is expected (see Supplementary Note \rom{13}). The remaining additional decoherence of the nitrogen spin is currently not understood.


The electron Ramsey experiment shows a beating frequency of 21.5(1) kHz (Fig. \ref{Figure4}d). As the data is an average over S1 and S2, this suggests an interaction between these two spins. Next, we will confirm this hypothesis and use the coupling between S1 and S2 to demonstrate an entangled state of two P1 centers.

{\bf{Entanglement of two dark electron spins}}

\noindent Thus far we have shown selective initialisation, control and single-shot readout of individual P1 centers within the bath. We now combine all these results to realize coherent interactions and entanglement between the electron spins of two P1 centers.

\begin{figure*}[tb]
\includegraphics[scale=0.9]{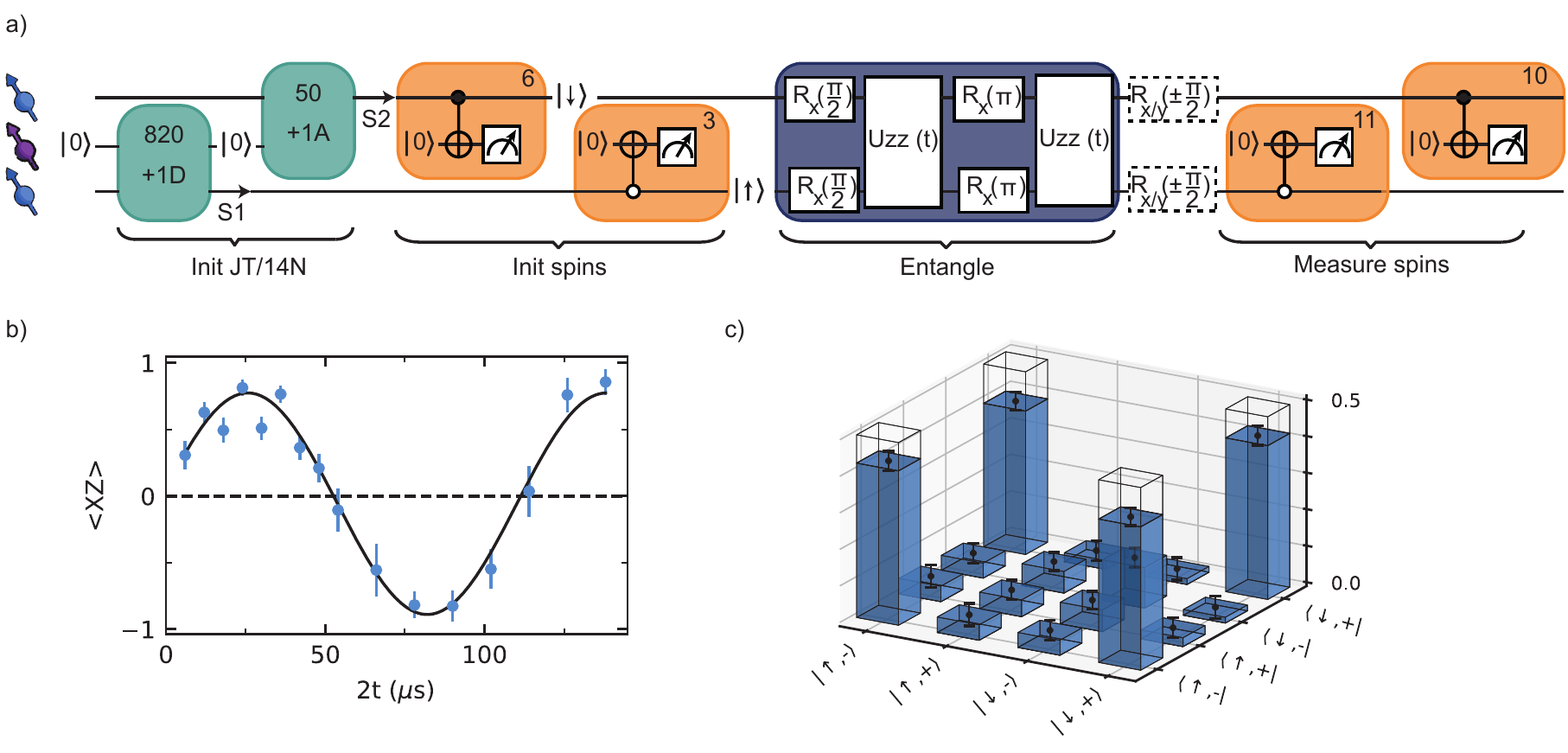}
\caption{\label{Figure5}
\textbf{Entanglement between two P1s.} a) Experimental sequence to measure coupling and generate entanglement between S1 and S2. DEER measurements initialize the JT axis and nitrogen state of S1 and S2 ($K$=820, 50 and $f$ = $f_{\mathrm{+1D}}$, $f_{\mathrm{+1A}}$), followed by DEER(y) measurements to initialize their electron spin states ($L$=6, 3). Two $\pi$/2 pulses and an evolution for time $2t$ under a double echo implements the $S_zS_z$ interaction with both spins in the equatorial plane of the Bloch sphere. This is followed by single qubit gates (dashed boxes) for full 2-qubit state tomography and two final DEER(y) measurements for electron spin readout. We apply an additional initial sequence ($K$ = 5, $f_{\mathrm{+1A}}$) to speed up the experiment (not shown in sequence, see Supplementary Note \rom{15}).  b) The coherent oscillation of the $\langle XZ \rangle$ as a function of interaction time $2t$ demonstrates a dipolar coupling $J$ = -$2\pi \cdot$17.8(5) kHz between S1 and S2. c) Density matrix of the S1 and S2 electron spins after applying the sequence as shown in (a) for $2t$ = $\pi$/$J$. The fidelity with the target state is $F$ = 0.81(5). Transparent bars indicate the density matrix for the target state $\ket{\Psi}$. All error bars indicate one statistical standard deviation.}
\end{figure*}

We first initialize both P1 centers (Fig. \ref{Figure5}a). We prepare S1 in the $\ket{+1,\mathrm{D}}$ state and S2 in the $\ket{+1,\mathrm{A}}$ state. By initializing the two P1 centers in these different states, we ensure that the spin transitions are strongly detuned, so that mutual flip-flops are suppressed and the interaction is effectively of the form $S_z S_z$. We then sequentially initialize both electron spins to obtain the initial state $\ket{\uparrow}_{S1}\ket{\downarrow}_{S2}$. As consecutive measurements can disturb the previously prepared degrees of freedom, the number of repetitions in each step is optimized for high total initialisation fidelity and success rate (Supplementary Note \rom{15}).

Next, we characterize the dipolar coupling $J$ between S1 and S2 (Fig. \ref{Figure5}b). We apply two $\pi$/2 pulses to prepare both spins in a superposition. We then apply simultaneous echo pulses on each spin. This double echo sequence decouples the spins from all P1s that are not in $\ket{+1, \mathrm{D}}$ or $\ket{+1,\mathrm{A}}$, as well as from the $^{13}$C nuclear spin bath and other noise sources. This way, the coherence of both spins is extended from $T_2^*$ to $T_2 $, while their mutual interaction is maintained. We determine the coupling $J$ by letting the spins evolve and measuring $\langle XZ \rangle$ as a function of the interaction time $2t$ through a consecutive measurement of both electron spins (Fig. \ref{Figure5}b). From this curve we extract a dipolar coupling $J$ = -$2\pi\ \cdot$\ 17.8(5) kHz between S1 in $\ket{+1,\mathrm{D}}$ and S2 in $\ket{+1,\mathrm{A}}$.

Finally, we create an entangled state of S1 and S2 using the sequence in Fig. \ref{Figure5}a. We set the interaction time $2t$ = $\pi$/$J$ so that a 2-qubit CPHASE gate is performed. The final state is (see Supplementary Note \rom{14}):
\begin{equation}
    \ket{\Psi} = \frac{\ket{\uparrow}_{S1}\ket{-}_{S2} + \ket{\downarrow}_{S1}\ket{+}_{S2}}{\sqrt{2}},
\end{equation}
with $\ket{\pm}$ = $\frac{\ket{\uparrow} \pm \ket{\downarrow}}{\sqrt{2}}$.  We then perform full 2-qubit state tomography and reconstruct the density matrix as shown in Fig. \ref{Figure5}c. The resulting state fidelity with the ideal state is $F$ = ($1$ + $\langle XZ \rangle$ - $\langle ZX \rangle$ - $\langle YY \rangle$)/4 =  0.81(4). The fact that $F>0.5$ is a witness for two-qubit entanglement \cite{OtfriedGuhneGezaEntanglementDetection}. The coherence time during the echo sequence ($\sim$ $700\ \upmu$s, see Methods) is long compared to $\pi$/$J$ ($\sim$ 28 $\upmu$s), and thus the dephasing during the 2-qubit gate is estimated to be at most 2$\%$. Therefore we expect the main sources of infidelity to be the final sequential single-shot readout of the two electron spin states – no readout correction is made – and the sequential initialisation of the two electron spins (Supplementary Note \rom{15}).

\noindent\textbf{DISCUSSION}

In conclusion, we have developed initialisation, control, single-shot readout, and entanglement of multiple individual P1 centers that are part of a bath surrounding an NV center. These results establish the P1 center as a promising qubit platform. Our methods to control individual dark spins can enable enhanced sensing schemes based on entanglement \cite{Vorobyov2020QuantumSensing, Cooper2019Environment-assistedDiamond, Goldstein2011Environment-AssistedMeasurement}, as well as electron spin chains for quantum computation architectures \cite{Yao2012ScalableProcessor, Ping2013PracticalityTechnologies, Yao2013QuantumRegisters,Schlipf2017AQubit}. Larger quantum registers might be formed by using P1 centers to control nearby $^{13}$C nuclear spins with recently developed quantum gates \cite{Bradley2019AMinute}. Such nuclear spin qubits are connected to the optically active defect only indirectly through the P1 electron spin, and could provide isolated robust quantum memories for quantum networks. Finally, these results create new opportunities to investigate the physics of decoherence, spin diffusion and Jahn-Teller dynamics \cite{Yang2016ProposalImprovin} in complex spin baths with control over the microscopic single-spin dynamics.


\noindent\textbf{METHODS}

\textbf{Sample.} We use a single nitrogen vacancy (NV) center in a homoepitaxially chemical-vapor-deposition (CVD) grown diamond with a $\langle 100 \rangle$ crystal orientation (Element Six). The diamond is isotopically purified to an approximate $0.01 \%$ abundance of $^{13}$C. The nitrogen concentration is $\sim$75 parts per billion, see Supplementary Note \rom{5}. To enhance the collection efficiency a solid-immersion lens was fabricated on top of the NV center \cite{Robledo2011High-fidelityRegister,Hadden2010StronglyLenses} and a single-layer aluminum-oxide anti-reflection coating was deposited \cite{Pfaff2014UnconditionalBits,Yeung2012Anti-reflectionDiamond}. 

\textbf{Setup.} The experiments are performed at $3.3$ Kelvin (Montana Cryostation) with the magnetic field $\vec{B}$ applied using three permanent magnets on motorized linear translation stages (UTS100PP) outside of the cryostat housing. We realize a long relaxation time for the NV electron spin ($T_1 >$ 30 s) in combination with fast NV spin operations (peak Rabi frequency $\sim$ 26 MHz) and readout/initialisation ($\sim$ 40 $\upmu$s/100 $\upmu$s), by minimizing noise and background from the microwave and optical controls \cite{Abobeih2018One-secondEnvironment}. Amplifier (AR 20S1G4) noise is suppressed by a fast microwave switch (TriQuint TGS2355-SM). Video leakage noise generated by the switch is filtered with a high pass filter. 

\textbf{Error analysis.} The data presented in this work is either a probability derived from the measurements, the mean of a distribution, or a quantity derived from those. For probabilities, a binomial error analysis is used, where $p$ is the probability and $\sigma = \sqrt{ p \cdot (1-p)/Q }$, $Q$ being the number of measured binary values. For the mean $\mu$ of a distribution, $\sigma_{\mu}$ is calculated as $\sigma/\sqrt{Q}$, where $\sigma$ is the square root average of the squared deviations from the mean and $Q$ is the number of measurements. Uncertainties on all quantities derived from a probability or a mean are calculated using error propagation.

\textbf{NV spin control and readout.} We use Hermite pulse envelopes \cite{VandersypenNMRComputation,Warren1984EffectsCoherence} to obtain effective microwave pulses without initialisation of the intrinsic $^{14}$N nuclear spin of the NV. We initialize and read out the NV electron spin through spin selective resonant excitation ($F$ = 0.850(5)) \cite{Robledo2011High-fidelityRegister}. Laser pulses are generated by acoustic optical modulators (637 nm Toptica DL Pro, for spin pumping and New Focus TLB-6704-P for single-shot spin readout) or by direct current modulation (515 nm laser, Cobolt MLD - for charge state control, and scrambling the P1 center state, see Supplementary Note \rom{7}). We place two modulators in series (Gooch and Housego Fibre Q) for an improved on/off ratio for the 637 nm lasers.

\textbf{Magnetic field stabilization.} During several of the experiments we actively stabilize the magnetic field via a feedback loop to one of the translation stages. The feedback signal is obtained from interleaved measurements of the NV $\ket{0} \leftrightarrow \ket{-1}$ transition frequency. We use the P1 bath as a three-axis magnetometer to verify the stability of the magnetic field during this protocol (see Supplementary Note \rom{11}), and find a magnetic field that is stable to $<$3 mG along $\bf \hat{z}$ and $<$20 mG along the $\bf \hat{x},\hat{y}$ directions.

\textbf{Optical reset of the P1 states.} Initialisation of the $^{14}$N, JT and charge states throughout the manuscript is achieved by heralded preparation. Therefore, the experimental rate depends on the relaxation rates for these states. We show that photoexcitation \cite{Heremans2009} of the P1s, causes relaxation and use this to speed up the experiments (see Supplementary Note \rom{7} for details). We apply a $\sim$5 $\upmu$s 515 nm laser pulse to ``reset'' the $^{14}$N, JT and charge states of P1 centers. Thereafter, we obtain information about the configuration of the P1 bath by a short DEER sequence with $K$=5 repetitions. From this information, we infer the likelihood for the desired configuration and determine whether to apply a new optical reset pulse or continue with the full initialisation sequence. This procedure speeds up the experiments with a factor $\sim$ 4.

\textbf{Initialisation and single-shot readout fidelity.} We define the combined initialisation and readout fidelity for S1 in $\ket{+1,\mathrm{D}}$ and S2, S3/S4 not in that state as

\begin{equation}
    F_{S1} = P(N(k+1)>N_{RO}|N(k)>N_{S1}),
\end{equation}

\noindent whereas for a mixture of all other possibilities we define

\begin{equation}
     F_{notS1} =  P(N(k+1)\leq N_{RO}|N(k)\leq N_{notS1}).
\end{equation}

\noindent In both cases $P(X|Y)$ is the probability to obtain X given Y. We then take the average fidelity of these two cases:

\begin{equation}
\label{eq:F_S1notS1}
    F = \frac{F_{S1} + F_{notS1}}{2}.
\end{equation}

We initialize and measure the electron spin state of P1 centers through a DEER(y) sequence following initialisation of the $\ket{+1,\mathrm{D}}$ state. Similarly, we use the correlation of consecutive measurements $M(k)$ and $M(k+1)$ to determine the combined initialisation and readout fidelity $F_{\ket{\uparrow}/\ket{\downarrow}}$. First, we define the fidelity for $\ket{\uparrow}$ as

\begin{equation}
    F_{\ket{\uparrow}} = P(M(k+1)>M_{RO}|M(k)>M_{\ket{\uparrow}}),
\end{equation}

\noindent and the fidelity for $\ket{\downarrow}$ as

\begin{equation}
     F_{\ket{\downarrow}} =  P(M(k+1)\leq M_{RO}|M(k)\leq M_{\ket{\downarrow}}).
\end{equation}

\noindent Finally, the average combined initialisation and readout fidelity is given as

\begin{equation}
    F_{\ket{\uparrow}/\ket{\downarrow}} = \frac{F_{\ket{\uparrow}} + F_{\ket{\downarrow}}}{2}.
\end{equation}

\noindent For a description of the optimization of the single-shot readout fidelities, we refer to Supplementary Note \rom{15}.

\textbf{Data analysis.}
The DEER measurements in Fig. \ref{Figure1}c are fitted to:
\begin{equation}
    a_0 + A_0\cdot \mathrm{Exp}[-(2\tau/T_{2,DEER})^2] \cdot (1 + B_0 \cos (\omega \cdot t/2))
\end{equation}

\noindent from which we find $T_{2,DEER}$ of 0.767(6), 0.756(7), 0.802(6) and 0.803(5) ms for $\ket{+1,\mathrm{A}}$, $\ket{+1,\mathrm{B}}$, $\ket{+1,\mathrm{C}}$ and $\ket{+1,\mathrm{D}}$, respectively. The obtained values for $\omega$ are 2$\pi \cdot$2.12(5), 2$\pi \cdot$2.14(3) and 2$\pi \cdot$2.78(6) kHz with corresponding amplitudes $B_0$ of 0.105(5), 0.218(7), and 0.073(4) for $\ket{+1,\mathrm{A}}$, $\ket{+1,B}$ and $\ket{+1,C}$, respectively. For $\ket{+1,\mathrm{D}}$ we fix $B_0$ = 0.

The DEER measurements with P1 initialisation (Fig. \ref{Figure3}a) and the P1 nitrogen Ramsey (Fig. \ref{Figure4}c) are fitted to:
\begin{equation}
\label{eq:coupling_singlespins}
A_1 \cdot e^{-(2\tau/T)^2}(\cos(\nu \cdot t/2)) + a_1.  
\end{equation}

\noindent For the dephasing time during the DEER sequence (here $t$ = 2$\tau$) we find $T$ = 0.893(5), 0.763(8) and 0.790(8) ms for S1, S2 and S3/S4 respectively. The obtained respective dipolar coupling constants $\nu$ are $2\pi \cdot$1.894(3), $2\pi \cdot$1.572(6) and $2\pi \cdot$1.001(6) kHz. For the P1 nitrogen Ramsey we find a dephasing time of $T$ = $T^*_{2N}$ = 0.201(9) ms.

Spin-echo experiments (Fig. 1c  and Fig. 5) are fitted to

\begin{equation}
    A_2\cdot e^{-(t/T)^n} + a_2.
    \label{eq:SpinEcho}
\end{equation}

\noindent For the NV spin-echo (Fig. \ref{Figure1}c), $T$ = $T_2$ = 0.992(4) ms with $n$ = 3.91(7). For the P1 nitrogen and electron (insets of Figure \ref{Figure4}c,d) $T$ is $T_{2 N}$ = 4.2(2) ms or $T_{2 e}$ = 1.00(4) ms with the exponents $n$ = 3.9(8) and $n$ = 3.1(5), respectively.  

The Ramsey signal for the P1 electron spin in Fig. \ref{Figure4}d is fitted to a sum of two frequencies with a Gaussian decay according to:

\begin{equation}
    a_3 + e^{-(t/T^*_{2, e})^2} \cdot \sum_{j=1}^2  ( A_j \cos ((f_{det}+(-1)^jf_b/2)t)+ \phi_j))/2,
    \label{eq:Ramsey_two_freq}
\end{equation}

\noindent which gives a beating frequency $f_b$ = $2\pi \cdot$21.5(5) kHz.  

The value $R$ (Fig. \ref{Figure3_partII}b) is defined as 

\begin{equation}
    R = \frac{P_{(+y)} - P_{(-y)}}{P_{(+y)} + P_{(-y)}},
    \label{eq:R_CouplingSign_methods}
\end{equation}

\noindent where $P_{(+y)}$ ($P_{(-y)}$) is the probability to read out the $^{14}$N spin in the $m_I = +1$ state when using a +y (-y) readout basis in the DEER(y) sequence used to initialize the electron spin (Fig. \ref{Figure3_partII}b, see Supplementary Note \rom{9}).

\textbf{Two-qubit gate fidelity.} We estimate the dephasing during the two-qubit CPHASE gate in Fig. \ref{Figure5} by extrapolation of the measured P1 electron $T_{2e}$ = 1.00(4) ms for a single spin-echo pulse (decoupled from all spins except those in $\ket{+1,\mathrm{D}}$). We use the scaling $T_2\propto 1/\sqrt{\langle n_{spins} \rangle}$ with $\langle n_{spins} \rangle$ the average number of spins coupled to during the measurement \cite{DeLange2012}. The two-qubit gate is implemented by a double echo and the two P1s are thus not decoupled from spins in $\ket{+1,\mathrm{D}}$ and $\ket{+1,\mathrm{A}}$, resulting in $T_2 \sim$ $T_{2e}/ \sqrt{2} \approx$ 700 $\upmu$s. Assuming the same decay curve as for $T_{2e}$ ($n=3.1$) this implies a loss of fidelity due to dephasing of $\sim$0.4$\%$. For a Gaussian decay ($n=2$) the infidelity would be $\sim$2$\%$.
\\
\\

\noindent\textbf{Acknowledgements}
We thank V.V. Dobrovitski, G. de Lange and R. Hanson for useful discussions. This work was supported by the Netherlands Organisation for Scientific Research (NWO/OCW) through a Vidi grant and as part of the Frontiers of Nanoscience (NanoFront) programme. This project has received funding from the European Research Council (ERC) under the European Union’s Horizon 2020 research and innovation programme (grant agreement No. 852410). This project (QIA) has received funding from the European Union’s Horizon 2020 research and innovation programme under grant agreement No 820445.


\noindent\textbf{Author contributions}
MJD, SJHL and THT devised the project and experiments. CEB, MJD, SJHL and HPB constructed the experimental apparatus. MJD and SJHL performed the experiments. MJD, SJHL, HPB, and THT analysed the data. ALM and MJD performed preliminary experiments. MM and DJT grew the diamond sample. MJD, SJHL and THT wrote the manuscript with input from all authors. THT supervised the project.

\noindent \textbf{Data availability} The datasets generated during and/or analysed during the current study are available from the corresponding author on reasonable request.

\noindent \textbf{Ethics declaration}
\noindent The authors declare no competing interests. 
\bibliographystyle{naturemag}
\bibliography{A_P1_main}

\clearpage

\end{document}



\widetext
\begin{center}
\textbf{\large Supplementary Information for \\``Entanglement of dark electron-nuclear spin defects in diamond''}
\end{center}

\renewcommand{\figurename}{\textbf{Supplementary Figure}}
\renewcommand{\tablename}{\textbf{Supplementary Table}}

\setcounter{equation}{0}
\setcounter{figure}{0}
\setcounter{table}{0}
\setcounter{page}{1}
\makeatletter
\renewcommand{\theequation}{S\arabic{equation}}
\renewcommand{\bibnumfmt}[1]{[S#1]}
\renewcommand{\citenumfont}[1]{S#1}

\tableofcontents{}
\clearpage

\section{SUPPLEMENTARY NOTE: System Hamiltonian}

\noindent In this note we describe the Hamiltonian of a system consisting of a P1 center and an NV center in diamond.

\subsection{P1 center}

\noindent The Hamiltonian of the P1 center is given by \cite{smith1959electron}:

\begin{equation}
\label{eq:S_Hp1}
    H_{\text{P1}_{\bf i}} = \gamma_e \vec{B}\cdot\vec{S} + \vec{S}\cdot {\bf \hat{A}_{ i}}\cdot \vec{I} + \gamma_n \vec{B}\cdot\vec{I} + \vec{I}\cdot {\bf \hat{P}_{ i}}\cdot \vec{I},
\end{equation}

\noindent where $\gamma_e$ ($\approx 2\pi \cdot 2.802$ MHz/G) and $\gamma_n $ ($\approx -2\pi \cdot0.3078 $ kHz/G) are the electron and $^{14}$N gyromagnetic ratio respectively. $\vec{B}$ is the external magnetic field vector and $\vec{S}$ and $\vec{I}$ are the electron spin-1/2 and nuclear spin-1 operator vectors. The P1 center exhibits a JT distortion along the axis of one of the four carbon-nitrogen bonds (denoted by $i \in$ $\{$A,B,C,D$\}$). This JT axis  defines the principal axis of the hyperfine interaction and the quadrupole interaction. 
\newline

\noindent The hyperfine tensor ($\bf \hat{A}_{ i}$) and quadrupole tensor ($\bf \hat{P}_{ i}$) can be obtained in any coordinate frame via a transformation of the diagonal hyperfine tensor, ${\bf A_{diag}}$ = diag[$A_x,A_y,A_z$] and diagonal quadrupole tensor, ${\bf P_{diag}}$ = diag[$P_x,P_y,P_z$]:

\begin{equation}
    {\bf \hat{A}_{ i}} = R^T \cdot {\bf A_{diag}} \cdot R
\end{equation}
and
\begin{equation}
    {\bf \hat{P}_{ i}} = R^T \cdot {\bf P_{diag}} \cdot R.
\end{equation}

\noindent Here $R$ is the rotation matrix  from the principal axis of the P1 center to any other coordinate frame defined by Euler angles $\{ \alpha, \beta, \gamma\}$: 
\begin{equation}
R(\alpha,\beta,\gamma) = \begin{pmatrix}

\cos{(\gamma)}\cos{(\beta)}\cos{(\alpha)} - \sin{(\gamma)}\sin{(\alpha)} & \cos{(\gamma)}\cos{(\beta)}\sin{(\alpha)} + \sin{(\gamma)}\cos{(\alpha)} & -\cos{(\gamma)}\sin{(\beta)} \\

-\sin{(\gamma)}\cos{(\beta)}\cos{(\alpha)}-\cos{(\gamma)}\sin{(\alpha)} & -\sin{(\gamma)}\cos{(\beta)}\sin{(\alpha)} + \cos{(\gamma)}\cos{(\alpha)} & \sin{(\gamma)}\sin{(\beta)} \\

\sin{(\beta)}\cos{(\alpha)} & \sin{(\beta)}\sin{(\alpha)} & \cos{(\beta)}

\end{pmatrix}.
\end{equation}

\noindent Due to the axial symmetry of the P1 center in its principal axis coordinate frame ($A_x$ = $A_y$ and $P_x = P_y$), this can be reduced to:

\begin{equation}
\label{eq:rot_mat}
R(\beta,\alpha) = \begin{pmatrix}

\cos{(\beta)}\cos{(\alpha)} & \cos{(\beta)}\sin{(\alpha)} & -\sin{(\beta)} \\

-\sin{(\alpha)} & \cos{(\alpha)} & 0 \\

\sin{(\beta)}\cos{(\alpha)} & \sin{(\beta)}\sin{(\alpha)} & \cos{(\beta)}

\end{pmatrix}.
\end{equation}


\noindent For any of the four JT axes, the hyperfine (quadrupole) tensor ${\bf \hat{A}_i}$ (${\bf \hat{P}_i}$) in the coordinate frame of the symmetry axis of the NV center is obtained via a transformation with angles $(\beta, \alpha)_i$, where $(\beta, \alpha)_i$ $\in$ $\{(109.5^{\circ},240^{\circ})_{\mathrm{A}},(109.5^{\circ},120^{\circ})_{\mathrm{B}},(109.5^{\circ},0^{\circ})_{\mathrm{C}},(0^{\circ},0^{\circ})_{\mathrm{D}}\}$.

\subsection{NV-P1 system}
\noindent The Hamiltonian of a coupled NV-P1 system (in the frame of the symmetry axis of the NV center) is given by the terms corresponding to the NV, the P1 and the terms describing their dipolar coupling:

\begin{equation}
\label{eq:Htot}
H_\text{tot} = H_\text{NV} + H_\text{P1} + H_\text{dipole}.
\end{equation}

\noindent For the NV we only consider the electron spin of the NV center:

\begin{equation}
    H_\text{NV} = \Delta J_z^2 + \gamma_e \vec{B}\cdot\vec{J}.
\end{equation}

\noindent where $\Delta$ = 2 $\pi\ \cdot$ 2.877 GHz is the zero-field splitting and $\vec{J}$ is the electron spin 1 vector. 

\noindent If we consider a point dipole coupling between the electron spins of the P1 center and the NV, separated by a vector $\vec{r}$, the dipole term can be written as:

\begin{equation}
H_\text{dipole} = \nu_{dip} \cdot \big(3(\vec{S}\cdot\hat{r})(\vec{J}\cdot\hat{r}) - \vec{S}\cdot\vec{J}\big),
\end{equation}

\noindent where $\nu_{dip} = \frac{-\mu_0 \gamma_e \gamma_e \hbar}{4 \pi r^3}$, $r = |\vec{r}|$ and $\hat{r} = \vec{r}/r$. Transforming to spherical coordinates using the definitions $r_x = r\sin(\theta)\cos(\varphi)$, $r_y = r\sin(\theta)\sin(\varphi)$ and $r_z = r\cos(\theta)$, gives

\begin{equation}\begin{split}
\label{eq:dipolar}
H_\text{dipole} &= \nu_{dip} \cdot \bigg[ S_x J_x \left(3\sin^2(\theta)\cos^2(\varphi)-1\right) + S_y J_y \left(3\sin^2(\theta)\sin^2(\varphi)-1\right) + S_z J_z \left(3\cos^2(\theta)-1\right) \\
&+ (S_x J_y + S_y J_x) 3\sin^2(\theta)\cos(\varphi)\sin(\varphi) + (S_x J_z + S_z J_x) 3\cos(\theta)\sin(\theta)\cos(\varphi) \\
&+ (S_y J_z + S_z J_y) 3\cos(\theta)\sin(\theta)\sin(\varphi)\bigg].
\end{split}\end{equation}

\noindent Due to the large difference between electron and $^{14}$N gyromagnetic ratio ($\gamma_e$/$\gamma_n \sim$ 9000) we expect the electron-nuclear NV-P1 dipolar coupling to be negligible and omit this term in the Hamiltonian. 

\subsection{Orientation in the crystal lattice}

\noindent In the diamond lattice, the P1 defect is located in a tetrahedral geometry with four surrounding carbon atoms. A single carbon atom can be positioned either directly above (orientation 1) or below (orientation 2) the P1 center's nitrogen atom, and the other three carbon atoms at 109.5$^{\circ}$ bond angles below or above respectively. 

If we consider the JT distortions along the $\bf \hat{z}$ axis (JT axis D) for these two orientations, the nitrogen atom either distorts in the -$\bf \hat{z}$ direction or the +$\bf \hat{z}$ direction. Therefore, these two orientations correspond to two spin Hamiltonians as in eq. \ref{eq:Htot}: $H_{tot,1}$ with $\bf \hat{A}_{D}$ = $\bf A_{diag}$ and $H_{tot,2}$ with $\bf \hat{A}_{D}$ = $R(180^{\circ},0^{\circ})^T \cdot$ $\bf A_{diag}$ $\cdot R(180^{\circ},0^{\circ})$ (similar for $\bf \hat{P}_{D}$). From eq. \ref{eq:rot_mat} it is evident that $R(180^{\circ},0^{\circ})^T \cdot$ $\bf A_{diag}$ $\cdot R(180^{\circ},0^{\circ})$ = $\bf A_{diag}$, and thus $H_{tot,1}$ = $H_{tot,2}$. Therefore, in the experiments performed in this work, we cannot distinguish between these two different orientations of P1 centers. 
 
\section{SUPPLEMENTARY NOTE: DEER and DEER(y) sequence}
 
\label{sec:DEER_sequences}
 
Here a more detailed description is given for the experimental sequences shown in Figs. 2b and 3b of the main text. We consider an idealized case of a single NV, a single P1 center and their magnetic dipolar coupling. In a diagonalized frame, the energy eigenstates and eigenvalues of the system are labelled as $\ket{m_s,m_{\uparrow/ \downarrow},m_I}$ and $\lambda_{m_s,m_{\uparrow/ \downarrow},m_I}$ respectively. The eigenstates form a 12-dimensional Hilbert space (the subspace m$_s$ = $\{$0,-1$\}$ of the NV, the electron spin 1/2 of the P1 center and its $^{14}$N spin 1).

\subsection{DEER}

We consider the sequence of unitary operations as applied during a DEER sequence with a single repetition $K$=1 (see Fig. 2b in the main text). First we apply $U_{NV,1}$ = $R_x$($\pi$/2)$_{NV} \otimes $ \(\mathbb{1}\)$_{P1}$, a $\pi$/2 rotation on the NV with phase $x$, followed by an evolution time $U_{evo}$($\tau$), which is given by diag[$e^{-i\lambda_{0,\uparrow,+1} \cdot \tau}$, $e^{-i\lambda_{0,\uparrow,0} \cdot \tau}$, ... ,$e^{-i\lambda_{m_s,m_{\uparrow/ \downarrow},m_I} \cdot \tau}$, ... ,$e^{-i\lambda_{-1,\downarrow,-1} \cdot \tau}$]. This is followed by a $\pi$ pulse on the NV; $U_{NV,2}$ = $R_x$($\pi$)$_{NV} \otimes $ \(\mathbb{1}\)$_{P1}$. Simultaneously, we apply a $\pi$ pulse on the P1 electron spin conditional on its $^{14}$N state $\ket{m_I}$ = $\ket{+1}$. This operation is described by $U_{P1}$ = \(\mathbb{1}\) $_{NV} \otimes R_x$($\pi$)$_{P1,e}\otimes\ket{+1}$ $\bra{+1}$ + \(\mathbb{1}\)$_{NV} \otimes$ \(\mathbb{1}\)$_{P1,e} \otimes \ket{0}$ $\bra{0}$ + \(\mathbb{1}\)$_{NV} \otimes$ \(\mathbb{1}\)$_{P1,e} \otimes \ket{-1}\bra{-1} $. Then there is another evolution time $U_{evo}$ ($\tau$) and finally a $\pi$/2 pulse $U_{NV,3}$ = $R_{-x}$($\pi$/2)$_{NV} \otimes $ \(\mathbb{1}\)$_{P1}$ with phase $-x$. We assume perfect $\pi$ and $\pi$/2 pulses between energy eigenstates. For an initial state $\rho_{1}$, we obtain a final state given by: 

\begin{equation}
    \rho_{f,1}(\tau) = U_{NV,3}U_{evo}(\tau)U_{P1}U_{NV,2}U_{evo}(\tau)U_{NV,1}\rho_{1}U_{NV,1}^{\dagger}U_{evo}(\tau)^{\dagger}U_{NV,2}^{\dagger}U_{P1}^{\dagger}U_{evo}(\tau)^{\dagger}U_{NV,3}^{\dagger}
\end{equation}

\noindent For the NV initialized in $\ket{0}$ but mixed P1 electron and $^{14}$N spin states, the initial density matrix is given as:

\begin{equation}
    \rho_{1} = \frac{1}{6} \ket{0}\bra{0}\otimes \mathbb{1} _{P1}
\end{equation}

\noindent The reduced density matrix of the NV as a function of interaction time is given by: 

\begin{equation}
    \rho_{NV}(\tau) = \sum_{m_{\uparrow,\downarrow}, m_I} \bra{m_{\uparrow,\downarrow}, m_I} \rho_{f,1}(\tau) \ket{m_{\uparrow,\downarrow}, m_I} = \frac{1}{6}\cos (\nu\cdot \tau)\big[\ket{-1}\bra{-1} - \ket{0}\bra{0}\big] + \frac{1}{6}\ket{0}\bra{0} + \frac{5}{6}\ket{-1}\bra{-1}.
\end{equation}

\noindent Here the effective NV-P1 dipolar coupling ($\nu$) is given as
\begin{equation}
\label{eq:effective}
\nu = \lambda_{-1,\downarrow,+1} - \lambda_{0,\downarrow,+1} - (\lambda_{-1,\uparrow,+1} - \lambda_{0,\uparrow,+1}).
\end{equation}
\noindent Note that at $\tau$ = $\pi$/$|\nu|$ we obtain the highest probability of measuring $\ket{0}_{NV}$. Upon measurement of $\ket{0}_{NV}$ we obtain the state: 

\begin{equation}
    \rho_{m_s = 0} = \frac{\ket{0}\bra{0}\rho_{f,1}(\tau)\ket{0}\bra{0}}{\Tr(\ket{0}\bra{0}\rho_{f,1}(\tau))} = \frac{1}{2} \ket{0}\bra{0} \otimes \mathbb{1} _{P1,e} \otimes \ket{+1}\bra{+1},
\end{equation}

\noindent and thus initialize the $^{14}$N state of the P1 center in $\ket{+1}$. 

\subsection{DEER(y)}

\noindent Now we consider DEER(y) (see main text Fig. 3b), which is sensitive to the P1 electron spin state. This sequence is the DEER sequence with the phase of the final $\pi$/2 pulse changed from $-x$ $\rightarrow$ $-y$, or $-x$ $\rightarrow$ $+y$. For an initial state $\rho_2$ and the phase of the final $\pi/2$ pulse as $-y$, we obtain the final state $\rho_{f,2}$ by:

\begin{equation}
    \rho_{f,2}(\tau) = U_{NV,4}U_{evo}(\tau)U_{P1}U_{NV,2}U_{evo}(\tau)U_{NV,1}\rho_{2}U_{NV,1}^{\dagger}U_{evo}(\tau)^{\dagger}U_{NV,2}^{\dagger}U_{P1}^{\dagger}U_{evo}(\tau)^{\dagger}U_{NV,4}^{\dagger}.
\end{equation}

\noindent Note that the final operator has been changed to $U_{NV,4}$ = $R_{-y}$($\pi$/2)$_{NV} \otimes $ \(\mathbb{1}\)$_{P1}$ with phase $-y$. If we consider the NV initialized in $\ket{0}$ and the $^{14}$N state of the P1 in $\ket{+1}$, the initial density matrix is given as:

\begin{equation}
    \rho_2 = \frac{1}{2} \ket{0}\bra{0} \otimes \mathbb{1} _{P1,e} \otimes \ket{+1}\bra{+1}. 
\end{equation}
 
\noindent The reduced density matrix of the NV as a function of interaction time is now given by:

\begin{equation}
    \rho_{NV}(\tau) = \sum_{m_{\uparrow,\downarrow}, m_I} \bra{m_{\uparrow,\downarrow}, m_I} \rho_{f,2}(\tau) \ket{m_{\uparrow,\downarrow}, m_I} = \frac{1}{2} \cos (\nu \cdot t) \big[ \ket{0}\bra{-1} + \ket{-1}\bra{0} \big] + \frac{1}{2} \ket{0}\bra{0} + \frac{1}{2} \ket{-1}\bra{-1}. 
\end{equation}

Upon measurement of $\ket{0}_{NV}$ we obtain the state: 

\begin{equation}
\rho_{m_s = 0} = \frac{\ket{0}\bra{0}\rho_{f,2}(\tau)\ket{0}\bra{0}}{\Tr(\ket{0}\bra{0}\rho_{f,2}(\tau))} =  \frac{1}{2} (1 + \sin (\nu \cdot \tau) ) \ket{0, \uparrow, +1}\bra{0, \uparrow, +1} + \frac{1}{2} (1 - \sin (\nu \cdot \tau) ) \ket{0, \downarrow, +1}\bra{0, \downarrow, +1}
\label{eq:rho_NV_DEERy}
\end{equation}

\noindent Note that at interaction time $\tau$ = $\pi$/(2$|\nu|$), the initialized state is either $\ket{0, \downarrow, +1}$ or $\ket{0, \uparrow, +1}$ depending on the sign of $\nu$. Applying a phase $+y$ instead of $-y$ in the final $\pi/2$ pulse effectively changes the signs in front of the $\sin(\nu \cdot \tau)$ terms in equation \eqref{eq:rho_NV_DEERy}, which is used in Fig. 4b.

\section{SUPPLEMENTARY NOTE: JT dependent coupling}
\label{sec:JT_dependent_coupling}

\begin{center}
\begin{figure*}[h]
\includegraphics[scale=1]{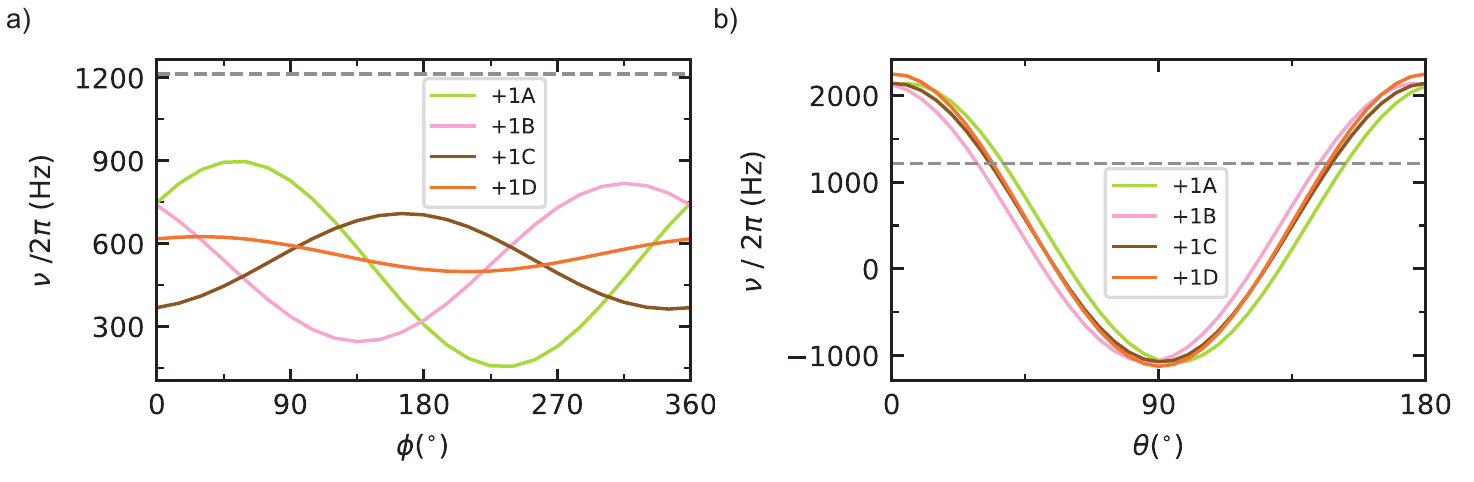}
\caption{\label{fig:JT_dependent} \textbf{Example showing JT dependent effective dipolar coupling $\boldsymbol{\nu}$.} \textbf{a)} Numerical simulation of $\nu$/$2\pi$ as a function of angle $\phi$ of the vector $\vec{r}$ between the NV and a P1. As an example we set $\theta$ = 45$^{\circ}$ and $|\vec{r}|$ = 35 nm. The magnetic field $\vec{B}$ and hyperfine/quadrupole parameters as obtained from the fitted DEER spectrum (main text Fig. 1b) are used. Colored curves indicate the P1 center in different states $\ket{+1,i}$. The dashed grey line indicates the value of $\nu_{dip}$/$2\pi$ as in eq. \ref{eq:dipolar}.  \textbf{b)} Similar to (a) but now as a function of $\theta$ ($\phi$ = 90$^{\circ}$).}
\end{figure*}
\end{center}

Mainly due to the anisotropic hyperfine interaction of the P1 center, the effective NV-P1 dipolar coupling $\nu$, which is the interaction observed in a DEER experiment (eq. \ref{eq:effective}), varies depending on the JT axis. We numerically calculate $\nu$ as a function of angles $\theta$ and $\phi$ of the vector $\vec{r}$ between an NV and a P1 using an exemplary case. Supplementary Fig. \ref{fig:JT_dependent}a, shows the calculated $\nu$/$2\pi$ for different states $\ket{m_I,i}$ as a function of angle $\phi$ as in eq. \ref{eq:dipolar} (here $\theta$ = 45$^{\circ}$). This figure demonstrates that the JT state can affect the effective dipolar coupling strength $\nu$ (eq. \ref{eq:effective}). For a P1 center in the states $\ket{+1,\mathrm{A}}$,$\ket{+1,\mathrm{B}}$ and $\ket{+1,\mathrm{C}}$ the principal axis of the hyperfine tensor is rotated 109.5$^{\circ}$ w.r.t. the $\bf{\hat{z}}$ axis (NV axis) and thus varying $\phi$ has a substantial effect on $\nu$. For a P1 center in $\ket{+1,\mathrm{D}}$ the hyperfine tensor's principal axis is along $\bf{\hat{z}}$ and therefore the dependence on $\phi$ is small. The observed dependence in the simulation for $\ket{+1,\mathrm{D}}$ is explained by the purposely slightly tilted magnetic field. The tilted magnetic field also explains the differences in maximal and minimal values of $\nu$ for the states $\ket{+1,\mathrm{A}}$,$\ket{+1,\mathrm{B}}$ and $\ket{+1,\mathrm{C}}$. The $\phi$ angles at which $\nu$ is at a maximum for these three states are shifted by $\Delta \phi \approx$ 120$^{\circ}$ w.r.t. each other as is expected from the three-fold rotational symmetry.

\FloatBarrier

Supplementary Fig. \ref{fig:JT_dependent}b illustrates that the relative differences of $\nu$ for different states are largest near the magic angles $\theta$ = $\pm$54.7$^{\circ}$. Note that information about the P1 positions can be obtained by combining the knowledge of which signal belongs to which P1 center for several $\ket{m_I,i}$ states (Fig. 2d main text) and measuring the NV-P1 dipolar coupling. This provides a future opportunity to determine the position of P1 centers (up to inversion symmetry) w.r.t. the NV.

\section{SUPPLEMENTARY NOTE: Fitting the Hamiltonian parameters}
\label{sec:SOM-fitting}

\begin{center}
\begin{figure*}[h]
\includegraphics[width = 0.97\textwidth]{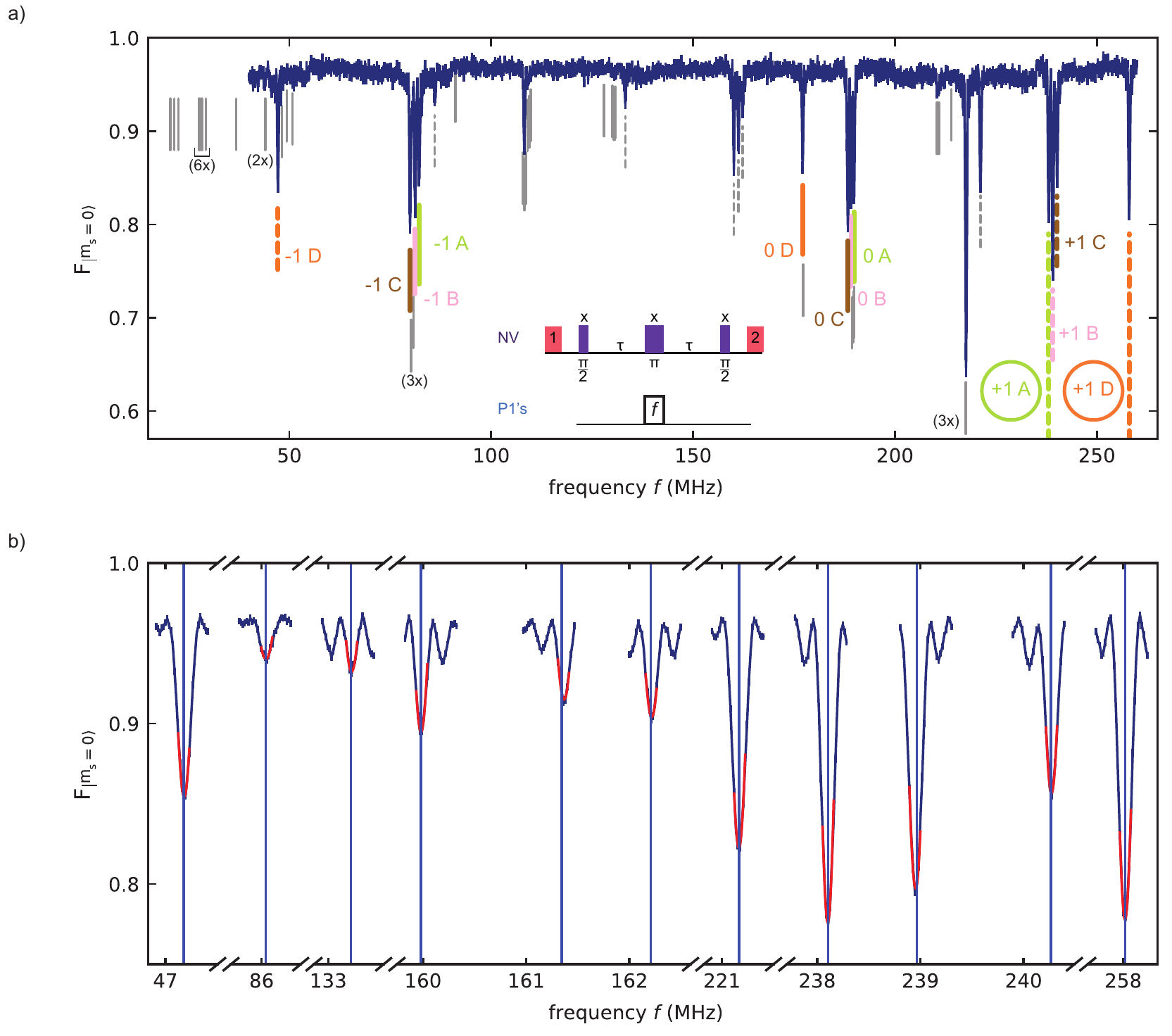}
\caption{\label{FigureS3a}
\textbf{Fitting the P1 Hamiltonian parameters.} a) Complete dataset corresponding to Fig. 1b (main text). No measurements were performed around the predicted transitions on the left. b) Measurement of 11 dips that are assigned to a well isolated transition (blue lines). Side lobes are due to Rabi oscillations. We fit a parabola (red) and extract the center frequency. Using a least squares fit of the center frequencies and the transitions, we extract the values of $\{$\textit{A{$_{\parallel}$}}, \textit{A{$_{\bot}$}}, \textit{P$_{\parallel}$}$\}$ and $\vec{B}$. Error bars indicate one statistical standard deviation, with a typical value $\num{2e-3}$, which is smaller than the data points.}
\end{figure*}
\end{center}

This section describes the fitting procedure used to obtain the Hamiltonian parameters $A_{\parallel}$, $A_{\bot}$, $P$, $B_x$, $B_y$ and $B_z$ from the DEER spectroscopy (see Fig. 1b and Supplementary Fig. \ref{FigureS3a}). We use parabolic fits of the measured dips to determine their center frequencies, and subsequently use a least squares method to minimize the difference between the measured frequencies and the frequencies resulting from diagonalization of the Hamiltonian (eq. \eqref{eq:S_Hp1}) for all 4 JT axes (see Supplementary Fig. \ref{FigureS3a}b). This method requires that we can assign which measured frequency (dips in Supplementary Fig. \ref{FigureS3a}a) belongs to which transition. We use a two step process. First, we take the four highest-energy transitions, which are well separated from any other transition, and use initial values for $A_{\parallel}$, $A_{\bot}$, $P$ \cite{Knowles2014} to obtain an estimate of the magnetic field vector. Second, we select 11 well-isolated transitions and corresponding experimental dips to fit $A_{\parallel}$, $A_{\bot}$, $P$ and the magnetic field vector.

To obtain an initial estimate of $B_x$, $B_y$ and $B_z$, we first perform a brute force optimization. We sweep $B_x$ and $B_y$ between $\pm$4 G and $B_z$ between 45 and 46 G in discrete steps and diagonalize the Hamiltonian (eq. \eqref{eq:S_Hp1}), where we use $\gamma_e$ = $2\pi \cdot$ 2.802495 MHz/G and $\gamma_n$ = - $2\pi \cdot$ 0.3078 kHz/G. For every combination we sort the four highest frequencies (corresponding to $\ket{+1,i}$) and calculate $\Delta = \sum_{i=1}^{4}(f_{i,\mathrm{exp}} - f_{i,\mathrm{theo}})^2$, where $f_{i,\mathrm{theo}}$ is the transition frequency obtained by diagonalization and $f_{i,\mathrm{exp}}$ is the measured frequency. We find a minimum in $\Delta$ for $\vec{B} = (2.53, 1.39, 45.37)^T$ G.

We then use 11 dips that we can unambiguously assign to a well-isolated transition and measure these dips with high accuracy, see Supplementary Fig. \ref{FigureS3a}b. We use this data to extract $A_{\parallel}$, $A_{\bot}$, $P$ and the magnetic field vector using a least squares fit. The fit obtains $\{$\textit{A{$_{\parallel}$}}, \textit{A{$_{\bot}$}}, \textit{P$_{\parallel}$}$\}$ = $\{$114.0264(9), 81.312(1), -3.9770(9)$\}$ MHz and $\vec{B}$ = $\{$ 2.437(2), 1.703(1), 45.5553(5)$\}$ G. The vertical lines in Supplementary Fig. \ref{FigureS3a}b are the $11$ transition frequencies calculated with these values. In order to provide a more quantitative comparison we provide the experimentally measured frequencies of the 11 dips in Supplementary Table \ref{tab:T2} alongside with their values calculated from equation \eqref{eq:S_Hp1}. Furthermore, by inspection, we identify 9 other dips in Supplementary Fig. \ref{FigureS3a}a. We determine their center frequency and compare them to their closest transition frequency resulting from the fitted parameters, see Supplementary Table \ref{tab:T2}.

\FloatBarrier

\begin{table}[h!]
  \centering
  \begin{tabular}{c|c|c|c|}
    Dip Nr & Experimental value (MHz) & Equation \eqref{eq:S_Hp1} (MHz) & $\ket{m_I, i}$ state \\ \hline
    1 & 47.183(1) & 47.179 & $\ket{-1, \mathrm{D}}$\\ \hline 
    2 & 86.042(2) & 86.040 & \\ \hline 
    3 & 133.227(2) & 133.219 & \\ \hline 
    4 & 159.9810(4) & 159.980 & \\ \hline 
    5 & 161.367(4) & 161.344 & \\ \hline 
    6 & 162.217(3) & 162.208 & \\ \hline 
    7 & 221.1641(7) & 221.1648 & \\ \hline 
    8 & 238.1027(6) & 238.1051 & $\ket{+1,\mathrm{A}}$\\ \hline 
    9 & 238.954(1) & 238.965 & $\ket{+1, \mathrm{B}}$\\ \hline 
    10 & 240.271(1) & 240.266 & $\ket{+1, \mathrm{C}}$\\ \hline 
    11 & 258.0176(6) & 258.018 & $\ket{+1, \mathrm{D}}$\\ \hline \hline
    12 & 79.898(4) & 79.961 & $\ket{-1, \mathrm{C}}$ \\ \hline
    13 & 81.16(2) & 81.10 & $\ket{-1, \mathrm{B}}$ \\ \hline
    14 & 82.106(8) & 82.119 & $\ket{-1,\mathrm{A}}$ \\ \hline
    15 & 108.18(2) & 108.20 &  \\ \hline
    16 & 177.1167(2) & 177.1476 & $\ket{0, \mathrm{D}}$ \\ \hline
    17 & 188.32(2) & 188.32 & $\ket{0,\mathrm{C}}$ \\ \hline
    18 & 189.15(6) & 189.11 & $\ket{0, \mathrm{B}}$ \\ \hline
    19 & 189.88(5) & 189.88 & $\ket{0, \mathrm{A}}$ \\ \hline
    20 & 217.5783(8) & 217.5709 &  \\ \hline

   \end{tabular}
   \caption{\label{tab:T2} Comparison of measured P1 transition frequencies (Supplementary Fig. \ref{FigureS3a}b) with closest values from equation \eqref{eq:S_Hp1}. We include the 11 frequencies of Supplementary Fig. \ref{FigureS3a}b used in the fitting, as well as 9 other identified dips in Supplementary Fig. \ref{FigureS3a}a, which is taken at a slightly different magnetic field compared to Supplementary Fig. \ref{FigureS3a}b.}
\end{table}
\FloatBarrier

\clearpage

\section{SUPPLEMENTARY NOTE: Estimate of P1 concentration}

\label{sec:P1concentration}

In this section we estimate the P1 concentration surrounding the NV center. We calculate an expected concentration ($C_d$) based on decoherence of the NV electron spin. In the secular approximation the dipolar interaction of a number ($N_d$) of P1 centers and the NV (see eq. \ref{eq:dipolar}) is given as:

\begin{equation}
    H_{int} = J_z \sum_{k = 1}^{N_d} 2\pi \cdot \nu_k \cdot (1-3\cos^2(\theta)) \cdot S_z = J_z \sum_{k = 1}^{N_d}  b_k \cdot S_z,
\end{equation}

\noindent where $2 \pi \cdot \nu_k  = \frac{-\mu_0 \gamma_e \gamma_e \hbar}{4 \pi r_k^3}$ and $\theta$ the polar angle in spherical coordinates. For a given concentration of P1 centers $C_d$, we consider a sphere around the NV center of radius $R$: 

\begin{equation}
    R = \bigg( \frac{3 \cdot V_{tot}}{4 \pi} \bigg)^{1/3}, 
\end{equation}

\noindent where $V_{tot} = \frac{N_d \cdot V_{unit}}{C_d \cdot 8}$ with $V_{unit}$ the volume of a diamond crystal unit cell, and 8 denotes the number of atoms within one unit cell. Within such a sphere we generate a number of P1 centers ($N_d$ = 40) at random positions and calculate $T_{2}^*$ due to a Gaussian spin bath \cite{de2010universal} as:

\begin{equation}
    T_{2}^* = \frac{\sqrt{2}}{\frac{1}{2}\sqrt{\sum_{k = 1}^{N_d}b_k^2}}.
\end{equation}

\noindent We repeat this procedure to generate $m$ = 10$^4$ different spatial configurations of P1 centers and calculate the average, $\langle T_{2}^* \rangle$. 

The result of this simulation as a function of P1 concentration $C_d$ is shown in Supplementary Fig. \ref{S14a}. For randomly positioned P1s, 1/$\langle T_{2}^* \rangle$ scales linearly with $C_d$. Note the difference in scaling with the average number of P1 centers coupled to during a DEER measurement, $\langle n_{spins} \rangle$, which scales as 1/$T_{2,DEER} \propto \sqrt{\langle n_{spins} \rangle}$ due to the fixed positions of the P1s \cite{DeLange2012}. We estimate the NV decoherence time due to coupling to P1s to be within two values (as indicated by the red vertical lines in Supplementary Fig. \ref{S14a}). The first value is the measured $T_{2,NV}^*$ = 97(3) $\upmu$s (1/$T_{2,NV}^* \approx$ 10 kHz). Here it is assumed that the P1 centers are the dominant spin bath and the effect of the $^{13}$C bath, and other magnetic field noise sources, on the NV decoherence is small. The second value is extrapolated from the measured $T_{2,DEER}$ = 0.803(5) ms for $\ket{+1,\mathrm{D}}$ and thus is given by $T_2^* \sim T_{2,DEER}/\sqrt{12} \approx$ 230 $\upmu$s (1/$T_2^* \approx$ 4 kHz). Here we assume approximately equal coupling strength in all 12 $\ket{m_I,i}$ states and an equal probability of occurrence for each state. The concentration is expected to be within these values and thus we estimate $C_d$ $\sim$ 75 ppb. 

\begin{figure*}[h]
\includegraphics[scale=1.3]{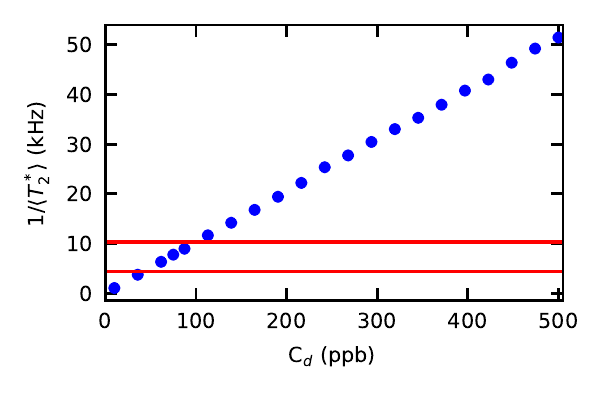}
\caption{\label{S14a} \textbf{Simulated $\boldsymbol{1/\langle T_2^* \rangle}$.} Simulation based on m = 10$^4$ different defect configurations of $N_d$ = 40 defects. The red vertical lines indicate estimated bounds for $1/\langle T_2^* \rangle$.}
\end{figure*}

\begin{figure*}[h]
\includegraphics[scale=0.9]{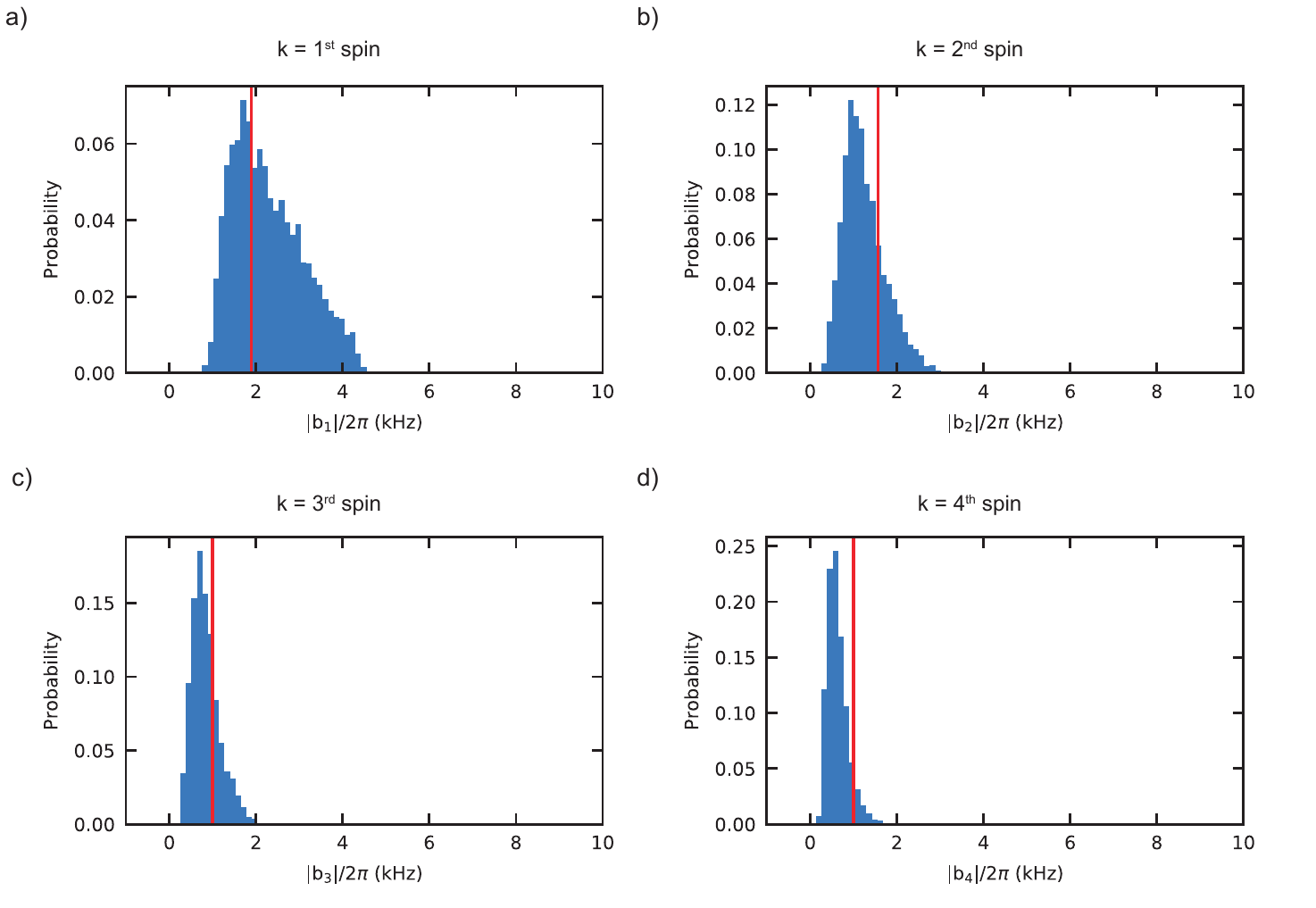}
\caption{\label{S14b} \textbf{Simulated distributions of the coupling strengths for the most strongly coupled P1 spins.} a),b),c),d) Distributions of $|$$b_k$$|$/$2\pi$ given 4 kHz $\leq$  1/$T_{2}^*$ $\leq$ 10 kHz and a concentration $C_d$ = 75 ppb.}
\end{figure*}

\FloatBarrier
Supplementary Fig. \ref{S14b} shows the probability density of $b_k$/$2\pi$ for the first four P1 spins, given a concentration $C_d$ = 75 ppb and the condition 4 kHz $\leq$  1/$T_{2}^*$ $\leq$ 10 kHz for the NV spin. These distributions demonstrate the expected values of $|b_k|$ to be close to the measured values as stated in Methods (indicated by the vertical red lines). The probability densities without a condition on 1/$T_{2}^*$ are shown in Supplementary Fig. \ref{S14c}. These plots show that given the concentration $C_d \sim$ 75 ppb, a large range of coupling strengths are possible for the nearest spin (including the measured value $|$17.8(5)$|$ kHz as measured between S1 and S2).

\begin{figure*}[h]
\includegraphics[scale=0.9]{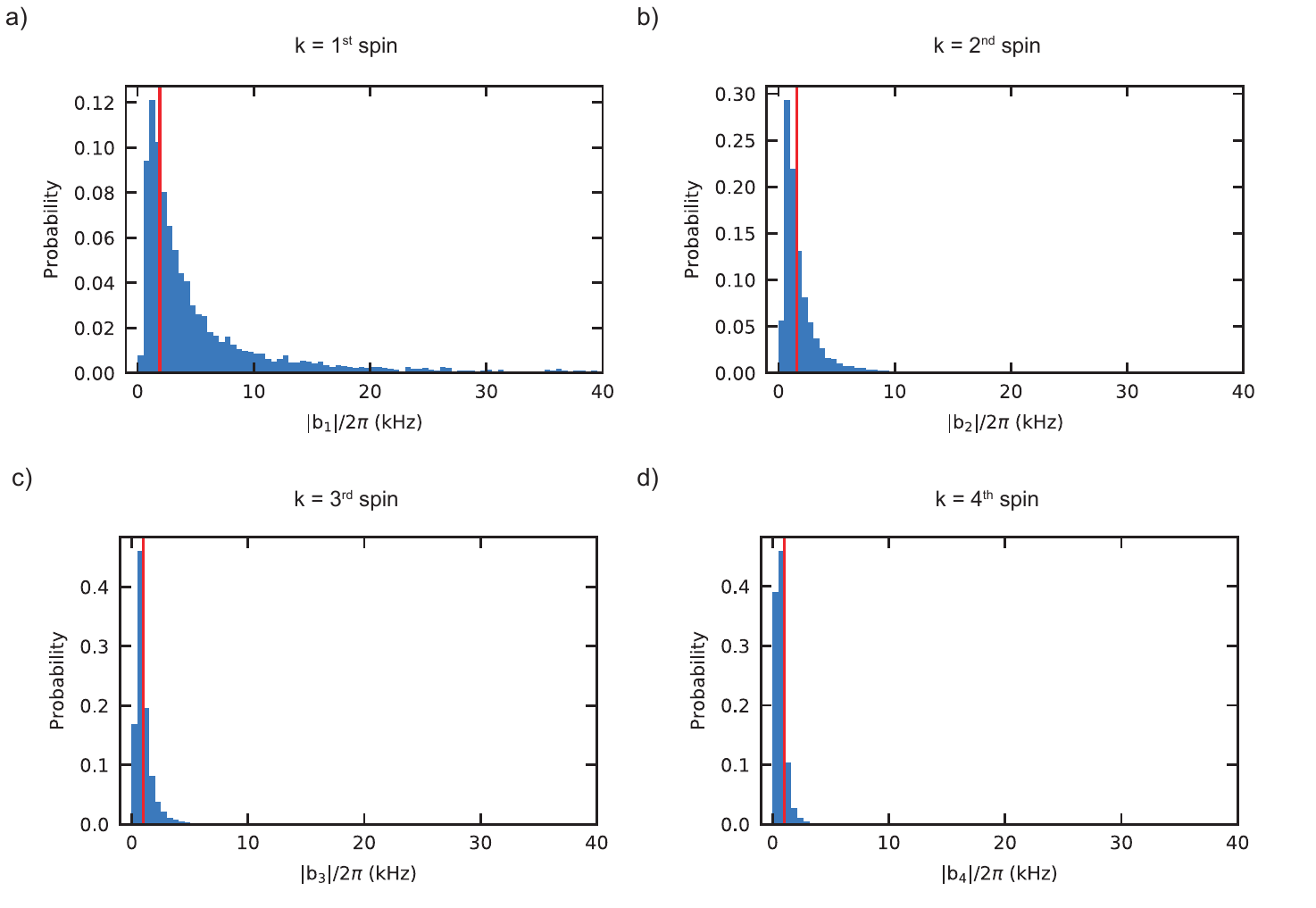}
\caption{\label{S14c} \textbf{Simulated distributions for the 4 most strongly coupled spins.} a),b),c),d) Distributions of $|b_k|$/$2\pi $ given a concentration $C_d$ = 75 ppb.}
\end{figure*}

\FloatBarrier

\section{SUPPLEMENTARY NOTE: Relaxation times of S1 during repetitive DEER measurements}
\label{sec:Relaxation_repRO}
\begin{figure*}[h]
\includegraphics[scale=1]{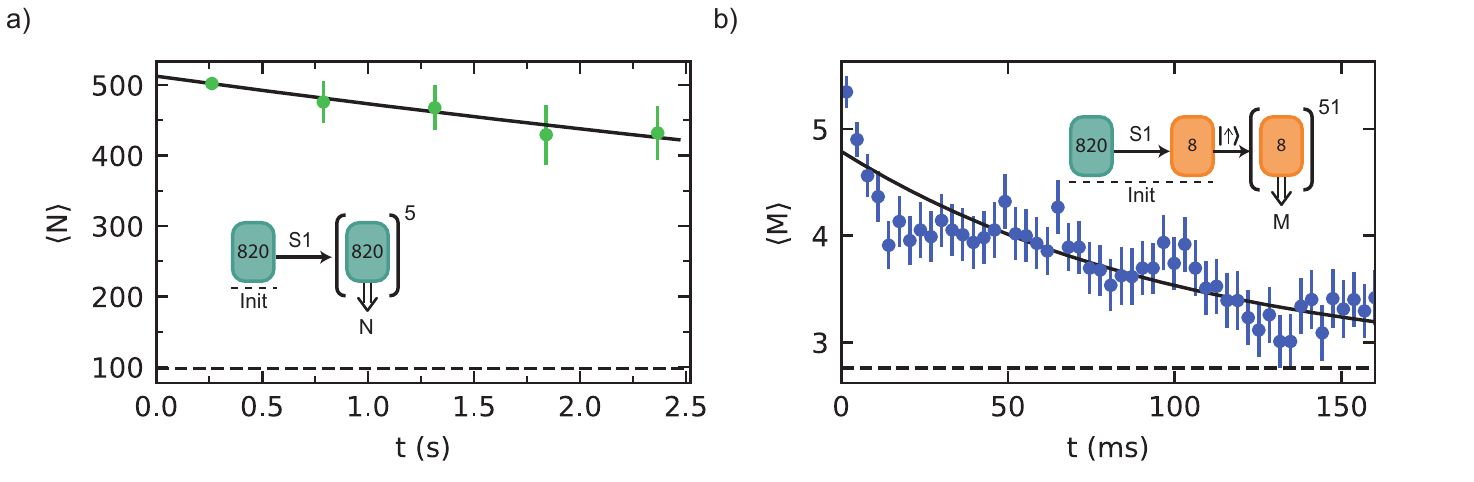}
\caption{\label{Fig:S16} \textbf{Relaxation during repetitive readout.} a) Relaxation from the $\ket{+1,\mathrm{D}}$ state while continuously performing DEER measurements. We fit (solid line) the curve to $o + A_0 \cdot e^{-t/T_{\ket{+1,\mathrm{D}}rep}}$, where $o$ is fixed to the uninitialized mean value (dashed line) and obtain $T_{\ket{+1,\mathrm{D}}rep}$ = 10(4) s. This timescale corresponds to $\sim$ 19 bins of DEER measurements with $K$=820 repetitions. b) Relaxation from the $\ket{\uparrow}$ state, we fit (solid line) the curve to $o_1 + A_1 \cdot e^{-t/T_{\ket{\uparrow}rep}}$, where $o_1$ is fixed to the uninitialized mean value (dashed line) and obtain $T_{\ket{\uparrow}rep}$ = 103(8) ms. This timescale corresponds to $\sim$ 32 bins of DEER(y) measurements with $L$=8 repetitions.}
\end{figure*}

This section characterizes the relaxation of the P1 $\ket{+1,\mathrm{D}}$ state (Supplementary Fig. \ref{Fig:S16}a) and the P1 electron spin in $\ket{+1,\mathrm{D}}$ (Supplementary Fig. \ref{Fig:S16}b) under repeated measurement sequences. We first prepare spin S1 in $\ket{+1,\mathrm{D}}$ using DEER measurements ($K$=820) and subsequently apply 5 sequential sets of DEER measurements with $K$=820. We plot the result of each set in Supplementary Fig. \ref{Fig:S16}a. We obtain a 1/e decay of $\sim$19 sets, showing the $\ket{+1,\mathrm{D}}$ state is stable over $\sim\num{1.5e4}$ DEER repetitions (including optical 637 nm pulses). 

To investigate the stability of the P1 electron state under repetitive DEER(y) measurements, we initialize S1 in $\ket{+1,\mathrm{D}}$ and prepare its electron state in $\ket{\uparrow}$ using $L$=8 DEER(y) measurements. Subsequently we apply 51 sequential sets of DEER(y) measurements with $L$=8 and plot the result of each set in Supplementary Fig. \ref{Fig:S16}b. We extract a 1/e decay for $\ket{\uparrow}$ of $\sim$32 sets, showing a stability over $\sim$250 DEER(y) repetitions. Comparing this with the 1/e decay of $\ket{+1,\mathrm{D}}$ shows that a single sequence is much more destructive for the P1 electron state than for the combined nitrogen and JT state. This limits the amount of DEER(y) sequences $L$ that can be used to initialize the P1 electron spin with high fidelity.

\FloatBarrier

\section{SUPPLEMENTARY NOTE: Preparation of P1 bath configurations by active optical reset}

\label{sec:scrambling}
\noindent Several of the experiments require initialisation of the charge, nitrogen and JT degrees of freedom for single P1 centers. To be able to distinguish the signals of the different P1 centers a large amount of measurement repetitions $K$ is required, which limits the experimental repetition rate. In this section we describe how we increase the experimental rate fourfold by dynamically resetting the P1 center states based on outcomes of early measurement repetitions. The key ideas are: (1) cases for which none of the P1 centers start in the desired state can be identified already after a few repetitions, so that the sequence can be aborted, and (2) the state of the P1 centers can be rapidly reset by a laser pulse (515 nm) before re-attempting the initialization.          


\subsection{Optimization to increase experimental rate}
In various experiments, we initialize the nitrogen- and JT-state of selected P1 centers by using the outcome of $K$ = 420 or $K$ = 820 DEER measurements to herald the desired state (Fig. 2a of the main text and Supplementary Fig. \ref{S5a}). In Supplementary Fig. \ref{figureS13b}a, we plot the histogram of $\sim$\num{3e5} DEER measurements in bins of $K$=420 on $\ket{+1,\mathrm{D}}$. In this dataset we define a successful (unsuccessful) initialisation of either S1 or S2 in $\ket{+1,\mathrm{D}}$ if the outcome $N$ of $K$=420 DEER measurements fulfils $N>180$ ($N\leq180$), see the green (red) part of the distribution in Supplementary Fig. \ref{figureS13b}a. Since $\sim$12\% of the distribution is in the green region, the success rate is limited. In Supplemenatry Fig. \ref{figureS13b}b, we inspect the measurement outcome $\mathrm{N1}$ of successful (green) and unsuccessful (red) initialisation attempts after the first $K$=$\Theta$ DEER sequences, where in this figure $\Theta = 5$. This shows that $\mathrm{N1}$ is low for many unsuccessful cases (red), indicating that the experiment can be sped up by aborting the sequence after a few repetitions if $\mathrm{N1}$ is below a threshold.  


We implement a Monte Carlo method to determine a good set of parameters to increase the experimental rate. A schematic of the method is depicted in Supplementary Fig. \ref{figureS13b}d. We sample from the dataset in Supplementary Fig. \ref{figureS13b}(a) and check after $\Theta$ DEER measurements (DEER 1) whether the outcome $\mathrm{N1}$ is below a threshold $\Lambda$. If so, we abort, sample again from the dataset and inspect DEER 1 again. If $\mathrm{N1} \geq\Lambda$, we continue until finishing 420 DEER measurements. If the outcome of the 420 DEER measurements is above 180, we accept it as a successful initialisation. We continuously sample from the dataset until we achieve 1000 successful runs ($\mathrm{N1}+\mathrm{N2}>180$).

We now sweep $\Theta$ and $\Lambda$ and calculate the average time $\langle \mathrm{T_{avg}} \rangle$ required to finish 1000 runs (see Supplementary Fig. \ref{figureS13b}d). Here we use that 1 DEER sequence takes 684 $\upmu$s and we introduce an overhead time of 1 ms for resetting the P1 center states (experimentally done with a laser pulse, see subsection below). Two critical parameters that determine $\langle \mathrm{T_{avg}} \rangle$ are 1) the probability $P(\mathrm{N1} \geq \Lambda)$ to pass the condition in DEER 1 and 2) the probability $P(\mathrm{N1}+\mathrm{N2}>180|\mathrm{N1} \geq \Lambda)$ to obtain $\mathrm{N1}+\mathrm{N2}>180$ conditioned on passing the condition in DEER 1. These probabilities can be extracted from Supplementary Fig. \ref{figureS13b}a and b and are shown in Supplementary Fig. \ref{figureS13b}c for the example case of $\Theta = 5$. From Supplementary Fig. \ref{figureS13b}d we extract that $\Theta = 5$ and $\Lambda = 3$ are parameters where $\langle \mathrm{T_{avg}} \rangle$ is small and we find an increase in experimental rate by a factor $\sim$ 5. This shows that the experiment can be sped up by aborting the sequence after 5 repetitions already if $\mathrm{N1}$ is below 3. Note, however, that this method assumes perfect randomization of the P1 center states by drawing random samples from the dataset.

\begin{center}
\begin{figure*}[h]
\includegraphics[scale=0.9]{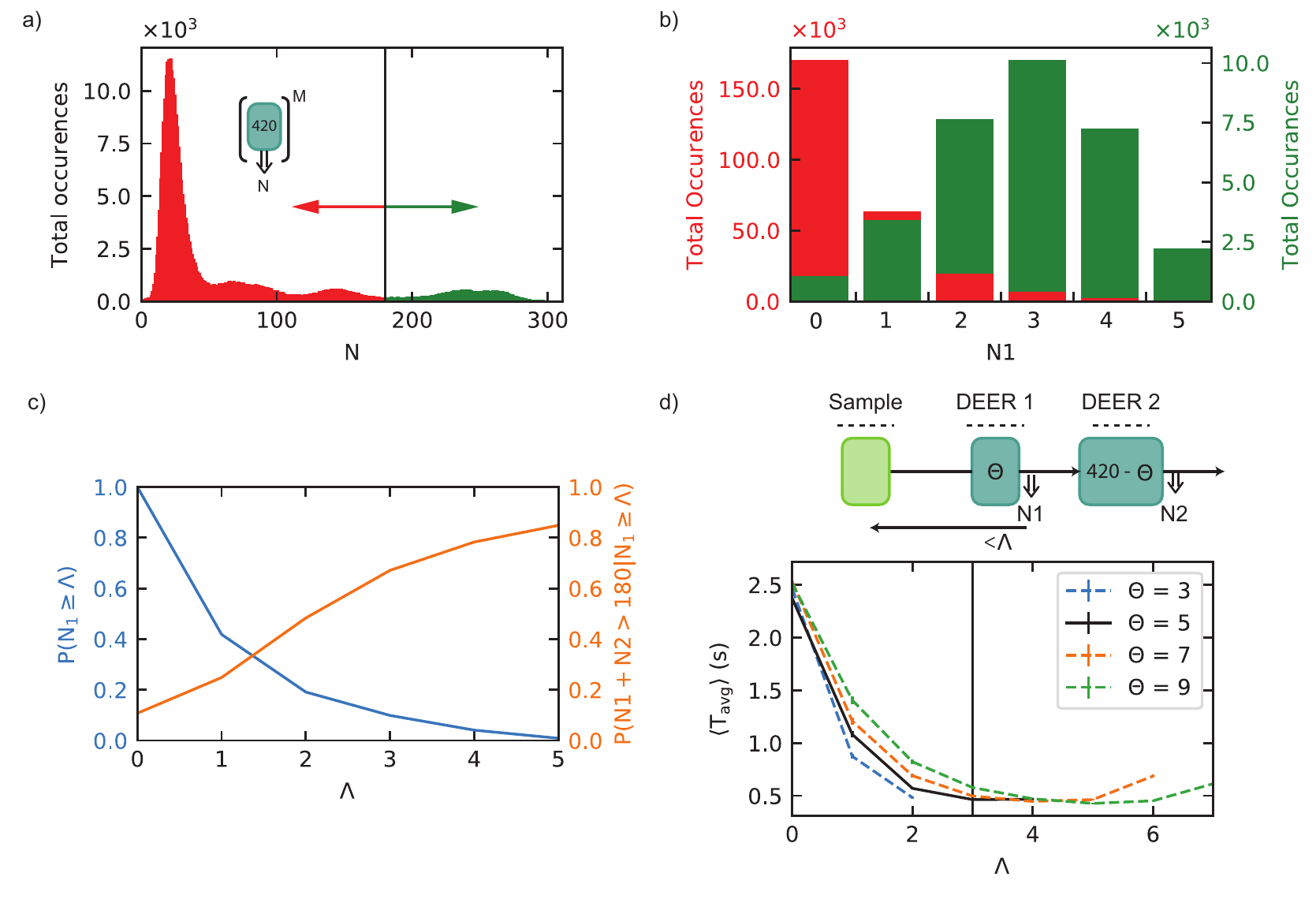}
\caption{\label{figureS13b} \textbf{Optimization to increase the experimental rate} a) The distribution of outcomes for $K=420$ DEER measurements. We define successful (unsuccessful) initializations as $N>180$ ($N\leq180$) indicated by the green (red). b) Distribution of outcomes after $\Theta$ DEER measurements, where as an example we take $\Theta$ = 5. Green (red) bars correspond to the green (red) datasets in (a). c) The probability $P(\mathrm{N1}\geq\Lambda)$ to measure an outcome $\mathrm{N1}\geq \Lambda$ after 5 DEER measurements (blue) and the probability $P(\mathrm{N1}+\mathrm{N2}>180|\mathrm{N1}\geq\Lambda)$ to obtain $N>180$ after 420 DEER measurements, conditioned on having $\mathrm{N1} \geq \Lambda$ in the first 5 DEER measurements. d) The average time ($\langle \mathrm{T_{avg}} \rangle$) to complete one successful run of the method shown on top. A run is successful if $\mathrm{N1}+\mathrm{N2}>180$. We calculate $\langle \mathrm{T_{avg}} \rangle$ as a function of $\Lambda$ for different settings of $\Theta$. We find a factor 5 increase in experimental rate for $\Theta$ = 5 and $\Lambda$ = 3 compared to no thresholding ($\Lambda = 0$).}
\end{figure*}
\end{center}
\FloatBarrier

\subsection{Active optical reset}
In this section we implement the method devised above to increase the experimental rate. We use photoexcitation \cite{Heremans2009} to efficiently randomize/reset the P1 center states after failed initialisation attempts. This method results in a fourfold increase of experimental rate. We observe a trade-off for the laser power, with higher laser power decreasing the rate due to spectral diffusion and ionization of the NV \cite{robledo2010control} and increasing the rate due to resetting the nitrogen and JT configuration of the P1 bath, and find an optimal working point.

\begin{center}
\begin{figure*}[h]
\includegraphics[scale=0.9]{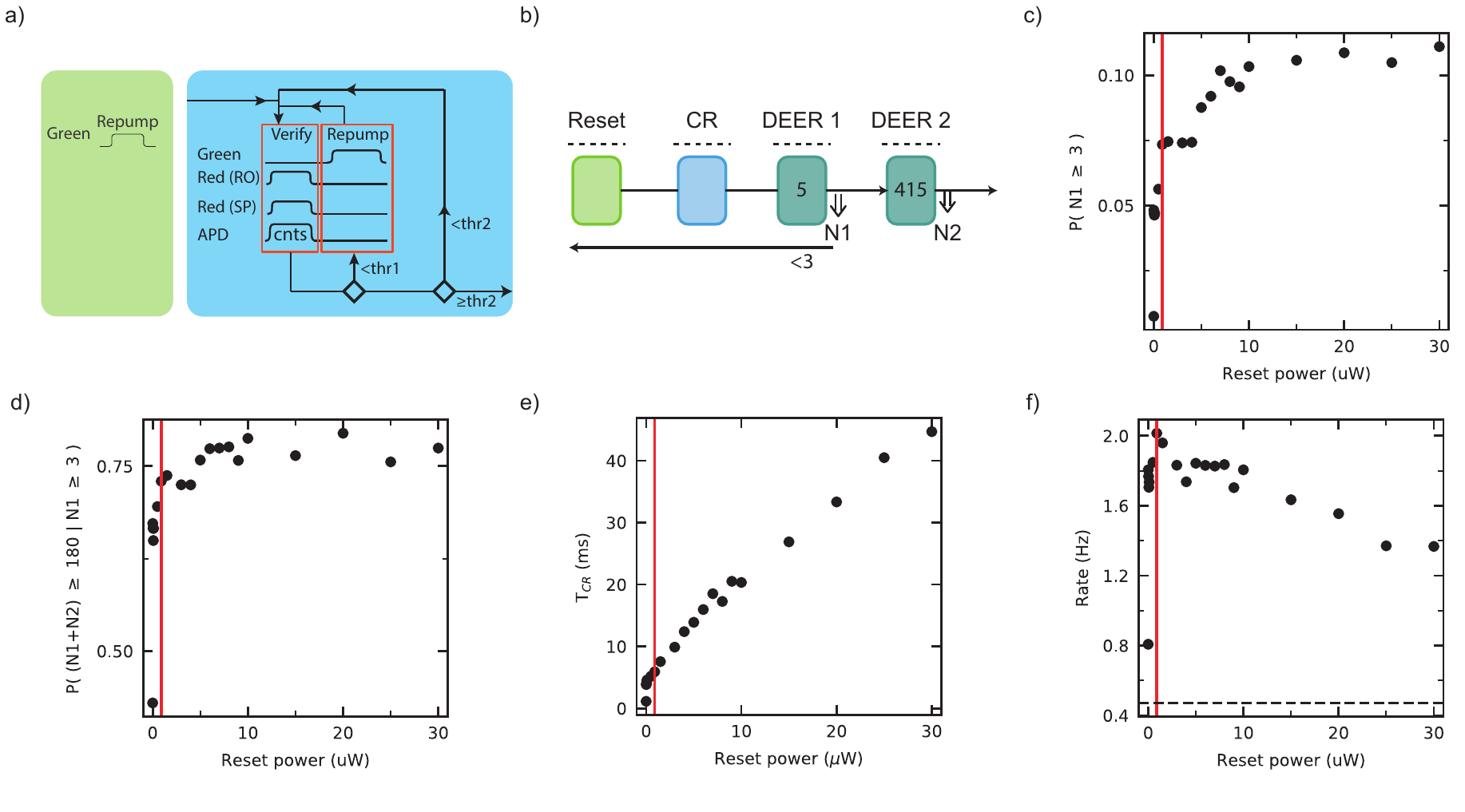}
\caption{\label{S13} \textbf{Active optical reset of P1 bath. } a) Illustration of the green (515 nm) optical reset pulse (left). Logical sequence showing the charge and resonance verification (CR) for the NV center (right). We count the number of photons (cnts) while resonantly exciting the NV center simultaneously at its readout (RO) and spin-pump (SP) transitions. If cnts $<$ thr1, the charge state is optically reset (repump, 30 $\upmu$W). The experimental sequence is continued if cnts $\geq$ thr2 (thr2 $>$ thr1). If thr1 $\leq$ cnts $<$ thr2, a new verification step is entered. b) Schematic of the experimental sequence. An initial  optical reset is performed (green, 5 $\upmu$s) to excite the P1 bath \cite{Heremans2009}. Thereafter, the CR scheme (blue, see (a)) is implemented until the experimental sequence is continued. This scheme is followed by a DEER sequence ($K$=5, see main text). Based on the outcome of DEER 1 we either continue to DEER 2 ($K$=415) or go back to the start of the experimental sequence to apply an optical reset pulse. c),d) Plots of the probabilities to pass the desired threshold after DEER 1 and DEER 2, as a function of the optical reset pulse power. e) Average duration of the CR scheme (T$_{CR}$) as a function of optical reset pulse power. f) Experimental rate given the desired threshold (N1 + N2) $\geq$ 180. The dashed horizontal line marks the rate without optical reset pulse and without feedback based on the outcome $\mathrm{N1}$. Vertical red lines in c),d),e),f) mark the reset power for maximal experimental rate.} 
\end{figure*}
\end{center}
\FloatBarrier

Our experiments require that the NV centre is in the correct negative charge state and that
its readout and spin-pump transitions are on resonance with the two 637 nm lasers. This is established by implementing a charge and resonance verification scheme (CR) \cite{Robledo2011High-fidelityRegister}. Additional to this scheme, we use an optical pulse to reset the P1 center states. Supplementary Figs. \ref{S13}a and b show the experimental sequence where the optical reset pulse is applied at the beginning. The CR scheme is thereafter implemented, followed by a short DEER sequence (DEER 1). At this point, feedback is implemented as the DEER 1 sequence provides information about the configuration of the P1 bath (see Supplementary Fig. \ref{figureS13b}b). Based on the method above, we apply an optical reset pulse if $\mathrm{N1}$ $<$ 3. In Supplementary Figs. \ref{S13}c,d), an increase of the probability of passing the desired thresholds for both DEER 1 and DEER 2 as a function of reset pulse power is shown. This demonstrates the reset pulse to be fully effective at a power of $\sim$10 $\mathrm{\upmu}$W. 

We observe an increase of the duration of the CR scheme as a function of reset power (Supplementary Fig. \ref{S13}e), likely due to spectral diffusion caused by photoexcitation of the P1 bath. The trade-off between optical reset and increased CR scheme duration becomes apparent in Supplementary Fig. \ref{S13}f, where the experimental rate is plotted as a function of optical reset power. The rate initially increases, followed by a decrease as the time required for CR verification becomes dominant. We find over a 4-fold increase of experimental rate at the optimal optical reset power of 0.9 $\mathrm{\upmu}$W. The observed increase in experimental rate here is close to theoretically predicted in the section above.
\\

\section{SUPPLEMENTARY NOTE: Correlation measurements for different JT axes}
\label{sec:JTAJTD_corr}

\noindent In this section we provide the background for the correlation measurement on different JT axes (Fig. 2d, main text), and include an analysis of the complete dataset (additional to the selected data in the main text). We derive the expected correlation value $C$ for a signal originating from a number $n$ of P1 centers in such correlation measurements. 

First, we define the regions of interest for the measurement outcomes in both JT axes ($N_{\ket{+1,\mathrm{A}}}$ and $N_{\ket{+1,\mathrm{D}}}$, see Supplementary Fig. \ref{S5a}). We fit the peaks in both distributions to Gaussian functions of the form $f = O + \sum_{q=1}^{q_{\mathrm{tot}}} A_q \cdot e^{-(N-N_{q,0})^2/2 \cdot \sigma_{q,0}^2}$, where $N$ corresponds to either $N_{\ket{+1,\mathrm{A}}}$ or $N_{\ket{+1,\mathrm{D}}}$ and $q_{\mathrm{tot}}$ is either 1 or 3. From these fits (black lines, Fig. \ref{S5a}), we obtain the FWHM of each Gaussian, and use these to define the areas of interest as shown in Supplementary Fig. \ref{S5b}. Note that these ranges for $N_{\ket{+1,\mathrm{D}}}$ are also used in experiments described in the main text for initialisation of S2 and S3/S4 in $\ket{+1,\mathrm{D}}$.

\begin{center}
\begin{figure*}[h]
\includegraphics[scale=0.9]{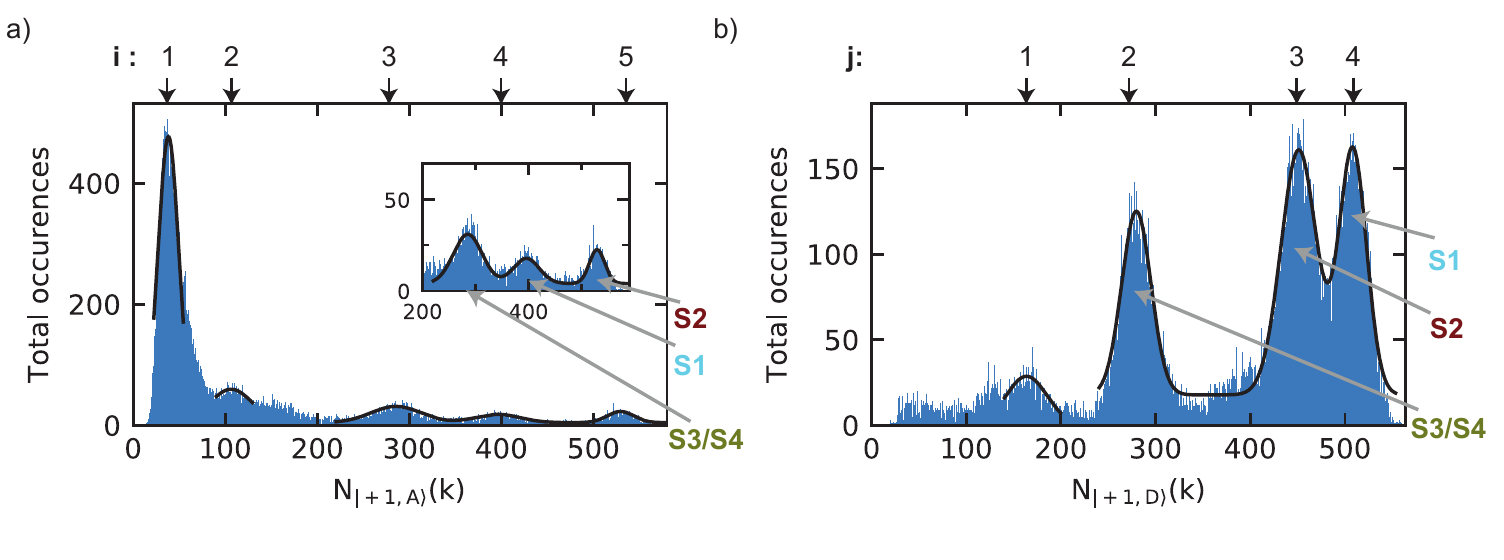}
\caption{\label{S5a} \textbf{Distributions of total occurrences. } a) All data from the measurement of Fig. 2d in the main text. The histogram shows the total occurrences plotted as a function of outcome N$_{\ket{+1,\mathrm{A}}}$ (irrespective of the outcome N$_{\ket{+1,\mathrm{D}}}$). The black solid line corresponds to the fit. Numbers corresponding to an index, indicate the center of each fitted Gaussian. Inset: enlarged view of the same histogram indicating S2, S1 and S3/S4. b) Similar to (a), but for measurements on $\ket{+1,\mathrm{D}}$. The data in (b) is taken with an increased rate using the active optical reset method described in Supplementary section \ref{sec:scrambling}, explaining why no peak at low counts is observed.} 
\end{figure*} 
\end{center}
\FloatBarrier

\begin{center}
\begin{figure*}[h]
\includegraphics[scale=0.9]{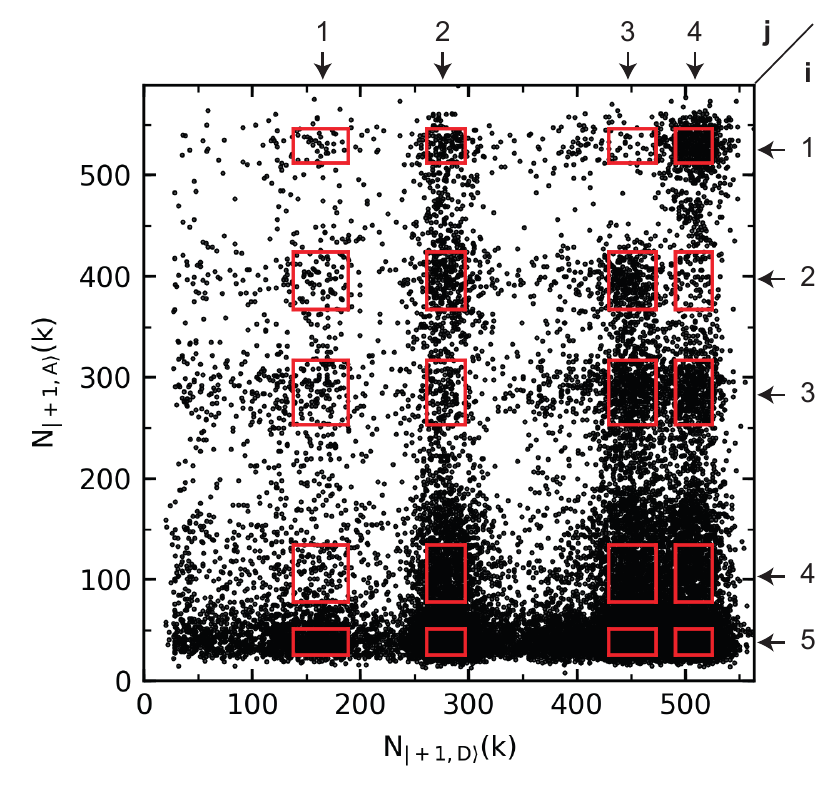}
\caption{\label{S5b} \textbf{Complete dataset corresponding to the measurement in Fig. 2d (main text).} The red rectangular areas illustrate the regions defined by the FWHM as obtained from the fitted curves in Supplementary Fig. \ref{S5a}. We denote each area by indices (i,j). The correlation values $C$ corresponding to these areas are shown in Supplementary Table \ref{tab:T1}.} 
\end{figure*}
\end{center}
\FloatBarrier

Second, we define the required probability functions. For outcome $N_{\ket{+1,\mathrm{D}}}$, the probabilities for obtaining $N^{min}_D \leq N_{\ket{+1,\mathrm{D}}} \leq N^{max}_D$ irrespective of $N_{\ket{+1,\mathrm{A}}}$ are given by:

\begin{equation}
    P(N^{min}_D \leq N_{\ket{+1,\mathrm{D}}} \leq N^{max}_D) =  P_D \big|_{min}^{max} = \int_{N^{min}_D}^{N^{max}_D} pdf_D \cdot dN_{\ket{+1,\mathrm{D}}},
\end{equation}

\noindent and, for outcome $N^{min}_A \leq N_{\ket{+1,\mathrm{A}}} \leq N^{max}_A$ irrespective of outcome $N_{\ket{+1,\mathrm{D}}}$, this is given as:

\begin{equation}
    P(N^{min}_A \leq N_{\ket{+1,\mathrm{A}}} \leq N^{max}_A) = P_A \big|_{min}^{max} = \int_{N^{min}_A}^{N^{max}_A} pdf_A \cdot dN_{\ket{+1,\mathrm{A}}}.
\end{equation}

\noindent $pdf_D$ and $pdf_A$ are probability density functions. We additionally define a conditional probability 

\begin{equation}
P( (N^{min}_A \leq N_{\ket{+1,\mathrm{A}}} \leq N^{max}_A)  |   (N^{min}_D \leq N_{\ket{+1,\mathrm{D}}} \leq N^{max}_D)) =\overline{P_A} \big|_{min}^{max} = \int_{N^{min}_A}^{N^{max}_A} \overline{pdf_A} \cdot dN_{\ket{+1,\mathrm{A}}}
\end{equation}

\noindent with $\overline{pdf_A}$ the probability density function for outcomes $N_{\ket{+1,\mathrm{A}}}$ given $N^{min}_D \leq N_{\ket{+1,\mathrm{D}}} \leq N^{max}_D$.



Finally, we derive a bound for the correlation function $C$. Consider $n$ P1 centers, each of which generates signal within $N^{min}_D \leq N_{\ket{+1,\mathrm{D}}} \leq N^{max}_D$ or within $N^{min}_A \leq N_{\ket{+1,\mathrm{A}}} \leq N^{max}_A$ if they are in the corresponding state. 
The correlation $C$ for consecutive measurements then satisfies: 



\begin{equation}
\label{eq:C}
    C = \frac{\overline{P_A} \big|_{min}^{max}}{P_A \big|_{min}^{max}} \geq \frac{n-1}{n}.
\end{equation}


The inequality in equation \ref{eq:C} is derived as follows. We consider $k$ P1 centers, each with an identical probability $p$ to be in any of the states $\ket{m_I,i}$ (here $p$ is approximately 1/12). We consider two signal regions $N^{min}_D \leq N_{\ket{+1,\mathrm{D}}} \leq N^{max}_D$  and $N^{min}_A \leq N_{\ket{+1,\mathrm{A}}} \leq N^{max}_A$, and assume that there are $n$ ($n\leq k$) P1 centers with a coupling that results in a signal in those regions. Note that  no signal is generated in the respective region if more than one P1 center is simultaneously in the state $\ket{+1,\mathrm{D}}$ or $\ket{+1,\mathrm{A}}$, as the NV then accumulates a different phase $\ddagger$.     
\vfill
\makebox[\linewidth]{\rule{\linewidth}{0.4pt}}{
\footnotesize{$\ddagger$ If two P1 centers with significant couplings to the NV center are both in the same state, the effective coupling is given by the sum or difference of the frequencies of the individual P1s, depending on the parity of their electron spins. However, because the P1 electron spin states relax rapidly under repeated measurement, an average over all spin states is observed for $K=820$ (Supplementary Fig. \ref{Fig:S16}b). Note that the probability that two or more of the P1s considered here are in the same state simultaneously is small.}} 

From the above, it follows that: 
\begin{equation}
P_D \big|_{min}^{max} = P_A \big|_{min}^{max} = n\cdot p \cdot (1-p)^{k-1}.    
\end{equation}
The observation of a signal in the region $N^{min}_D \leq N_{\ket{+1,\mathrm{D}}} \leq N^{max}_D$ means that one of the $n$ spins is in state $\ket{+1,\mathrm{D}}$ and thus conditional probability for the second measurement becomes:
\begin{equation}
    \overline{P_A} \big|_{min}^{max} = (n-1) \cdot p \cdot (1-p)^{k-2} = \frac{n-1}{n}\cdot \frac{1}{1-p} \cdot P_A \big|_{min}^{max} \geq \frac{n-1}{n} \cdot P_A \big|_{min}^{max},
\end{equation}
which yields equation \ref{eq:C}. In a similar way it follows that $\overline{P_A} \big|_{min}^{max} \geq P_A \big|_{min}^{max} \geq \frac{n-1}{n} P_A \big|_{min}^{max}$, if the signal $N^{min}_A \leq N_{\ket{+1,\mathrm{A}}} \leq N^{max}_A$ is generated by $m$  spins that are not necessarily the same as the $n$ spins that generate signal $N^{min}_D \leq N_{\ket{+1,\mathrm{D}}} \leq N^{max}_D$.


The values of $C$ obtained for all areas as indicated in Supplementary Fig. \ref{S5b} (red rectangles) are shown in Supplementary Table \ref{tab:T1}. For the areas (i,j) = (1,3) and (2,4) (as in Fig. 2d in the main text) we obtain $C$ = 0.22(4) and 0.40(5) respectively, which indicates a single spin. For (i,j) = (3,4) (as in Fig. 2d main text), $C$ = 0.47(4) indicating 1 or 2 spins. For areas such as (i,j) = (5,1), (5,2), (5,3) and (5,4)  we find C $\approx$ 1 indicating a large number of spins $n$. For these areas one would indeed expect a larger number of spins because they correspond to weaker dipolar coupling to the NV. Interestingly, some areas have a $C$ value that is significantly above unity such as (1,4), (2,1) and (2,2). For the area (1,4) the highest value is observed ($C$ = 2.2(2)), this suggests that there might be preferred combinations of $\ket{m_i,i}$ states for $S1$ and $S2$.  

\begin{table}[h!]
  \centering
  \begin{tabular}{c|c|c|c|c|}
    i\textbackslash j & 1 & 2 & 3 & 4 \\ \hline
            1 & 1.0(2) & 1.0(1) & 0.22(4) & 2.2(2)  \\ \hline
            2 & 1.6(2) & 1.5(1) & 1.20(8) & 0.40(5)    \\ \hline
            3 & 1.4(1) & 0.47(4)  & 1.14(6) & 1.15(6)  \\ \hline
            4 & 0.74(7) & 1.17(5) & 1.0(1) & 0.94(4)   \\ \hline
            5 & 0.99(4) & 0.94(2) & 1.00(2) & 0.97(2)   \\ \hline

   \end{tabular}
\captionsetup{justification=centering}
   \caption{  \label{tab:T1} $C$ values for areas as shown in Supplementary Fig. \ref{S5b}}
\end{table}
\FloatBarrier


\section{SUPPLEMENTARY NOTE: NV-P1 dipolar coupling sign  in $\ket{+1,\mathrm{A}}$}
\label{sec:couplingsign}
\normalsize

This section demonstrates coherent control of the $^{14}$N spin and extracts the sign of the NV-P1 dipolar coupling for the electron spin of S1, S2 and S3/S4 in the $\ket{+1,\mathrm{A}}$ state (similar as in Fig. 4b of the main text for $\ket{+1, \mathrm{D}}$). First, we choose an initialisation sequence with a low $K$ ($K$=6), which with high probability prepares either S1, S2 or S3/S4 in $\ket{+1,\mathrm{A}}$. Then we calibrate a $\pi$ pulse that implements a CNOT between the electron (control) and nitrogen (target) for a P1 in the $\ket{+1A}$ state, see Supplementary Fig. \ref{FigureS4}a. This pulse is only resonant with the $m_I = +1 \leftrightarrow 0$ transition, if the P1 electron spin is in the $\ket{\uparrow}$ state.




We use the same experimental procedure as in the main text for $\ket{+1, \mathrm{D}}$, to determine the NV-P1 coupling signs (Supplementary Fig. \ref{FigureS4}b). The result shows that spin S1 and S3/S4 have a positive coupling sign, whereas it is negative for S2. Note that S1, S2 and S3/S4 have the same NV coupling sign in $\ket{+1,\mathrm{A}}$ and $\ket{+1,\mathrm{D}}$.

\begin{center}
\begin{figure*}[ht]
\includegraphics[width = 0.95\textwidth]{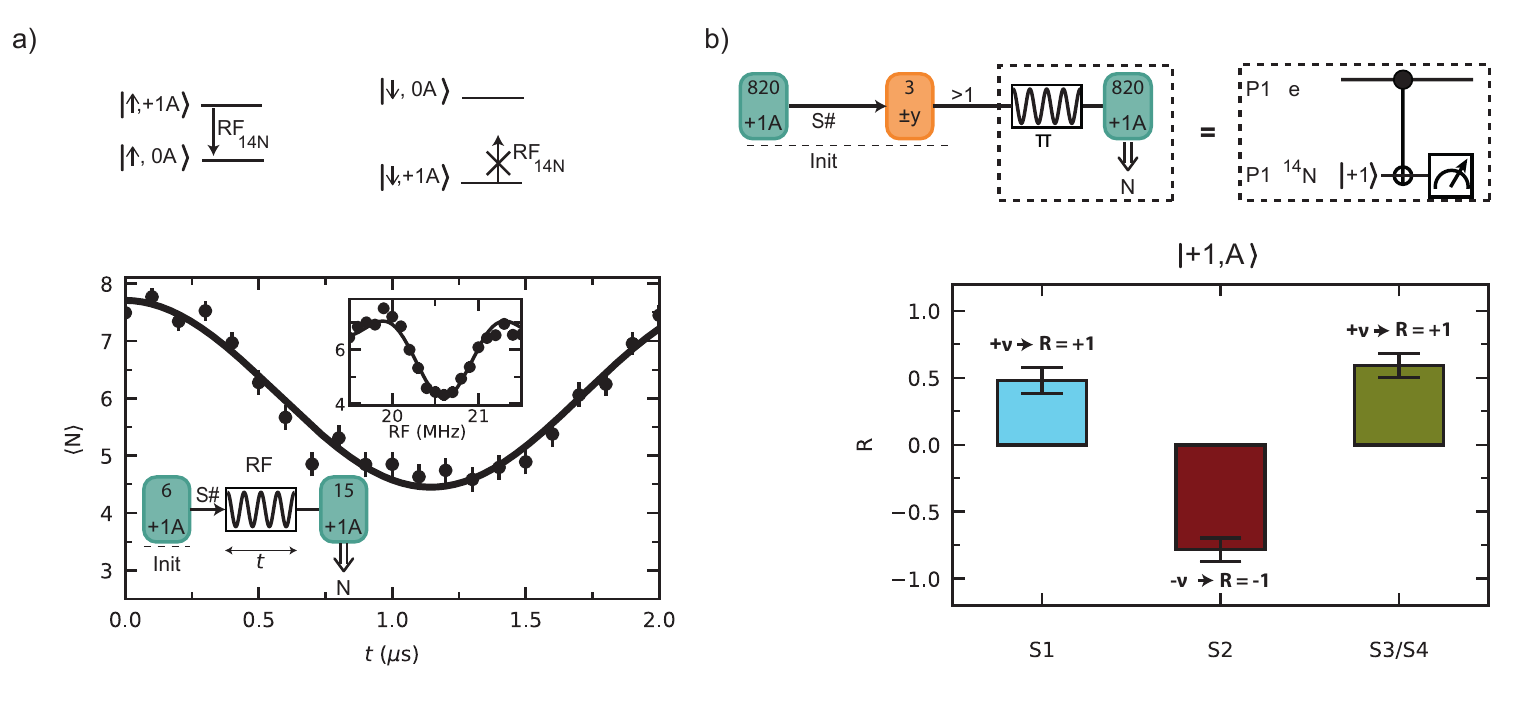}
\caption{\label{FigureS4}
\textbf{Coupling sign of NV-P1 dipolar interaction for different P1 centers in the $\boldsymbol{\ket{+1,\mathrm{A}}}$ state.} a) Top: energy level structure of the P1 electron spin in the $\{ \mathrm{0A, +1A} \}$ subspace. Bottom: A short initialisation sequence ($K$=6) prepares either S1, S2 or S3/S4 in $\ket{+1,\mathrm{A}}$ (without knowledge of which P1 is prepared in which run) and the length $t$ of a pulse at frequency $RF = RF_{14N}$ = 20.55 MHz is varied. The nitrogen spin is driven conditionally on the electron spin state (top). Inset: NMR spectrum obtained by varying the frequency ($RF$) for a fixed pulse duration $t$. b) Value of $R$ (see Methods) measured for the P1s in $\ket{+1,\mathrm{A}}$. Positive values correspond to a positive coupling sign.}
\end{figure*}
\end{center}
\FloatBarrier

\section{SUPPLEMENTARY NOTE: NV-P1 coupling in $\ket{+1,\mathrm{A}}$}
In Fig. 3a of the main text the effective NV-P1 dipolar coupling (equation \eqref{eq:effective}) for S1, S2, S3/S4 in $\ket{+1,\mathrm{D}}$ is measured. Here, we measure the effective dipolar coupling of these spins in $\ket{+1,\mathrm{A}}$, see Supplementary Fig. \ref{FigureS6}. We initialize each P1 center by setting the requirement that the outcome of the DEER initialisation sequence is within the range for $\ket{+1,\mathrm{A}}$ (see Supplementary Fig. \ref{S5a}a).

We obtain effective dipolar couplings $\nu$ of 2$\pi\ $ $\cdot$ 1.35(2), 2$\pi\ $ $\cdot$ 2.006(9) and 2$\pi\ $ $\cdot$ 1.06(2) kHz for S1, S2 and S3/S4 respectively, using the fit function in eq. 11 (see Methods). A comparison with the couplings for $\ket{+1,\mathrm{D}}$ (Fig. 3a in the main text) shows that the effective NV-P1 dipolar coupling strength differs for the two JT axes (see section \ref{sec:JT_dependent_coupling} for the theoretical treatment).

\begin{center}
\begin{figure*}[h]
\includegraphics[width = 0.5\textwidth]{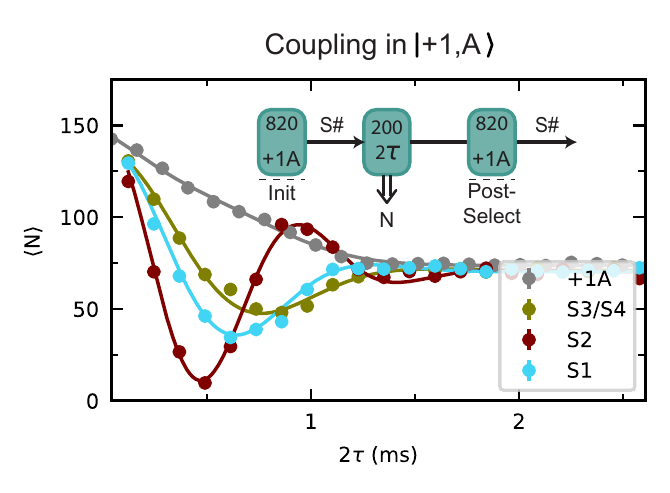}
\caption{\label{FigureS6}
\textbf{Measuring NV-P1 coupling strength in $\boldsymbol{\ket{+1,\mathrm{A}}}$.} We initialize S1, S2 or S3/S4 in $\ket{+1,\mathrm{A}}$ and vary the interaction time 2$\tau$ of a DEER sequence ($K$=200). To improve the signal, the results are post-selected on again obtaining $\ket{+1,\mathrm{A}}$. Inset: experimental sequence. Grey: without P1 initialisation (data from Fig. 1c main text).}
\end{figure*}
\end{center}
\FloatBarrier

\section{SUPPLEMENTARY NOTE: Magnetic field stability}
\label{sec:magnet}

Here we describe the magnetic field stabilisation used during the Ramsey and entanglement experiments in this manuscript, as well as during the data in Supplementary Fig. \ref{FigureS3a}b. We use three permanent magnets on motorized linear translation stages to create a static magnetic field. To compensate for magnetic field fluctuations we repeatedly measure the NV $\ket{0} \leftrightarrow \ket{-1}$ frequency ($f_{NV}$) and this signal is fed back to one of the magnet stages until ($f_{t}$ - 1.5 kHz $\leq$ $f_{NV}$ $\leq$ $f_{t}$ + 1.5 kHz), with target frequency $f_{t} = 2.749692$ GHz. We interleave this protocol (typically every 20 minutes) with repetitions of an experimental sequence.

\begin{center}
\begin{figure*}[h]
\includegraphics[width = 0.97\textwidth]{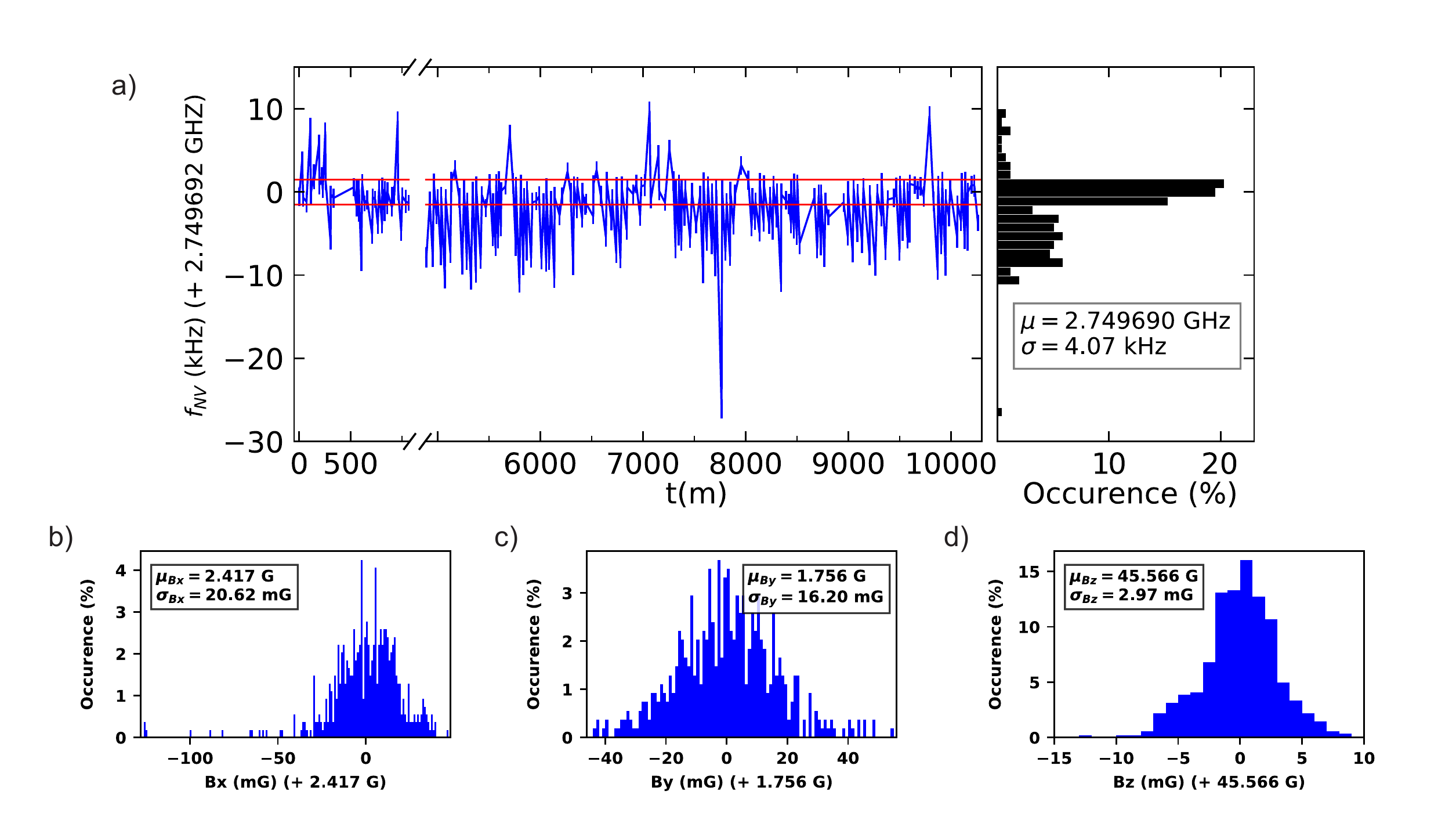}
\caption{\label{FigureS7}
\textbf{Magnetic field stability.} (a) Time trace of all measured NV resonance frequencies before and after performing repetitions of an experimental sequence. The red lines indicate the region ($\pm$ 1.5 kHz) to within which we stabilize before each measurement. (b,c,d) Distribution of measured $B_x$, $B_y$ and $B_z$ during 220 hours, while stabilizing $f_{NV}$.}
\end{figure*}
\end{center}

Supplementary Fig. \ref{FigureS7}a shows the time trace of $f_{NV}$ before and after interleaved experiments. The two red lines indicate $f_{t} \pm 1.5$ kHz. The magnetic field can freely drift during the experimental sequences, but right before starting the next run the field is stabilized. From the histogram we infer that on average $f_{NV}$ fluctuates with $\sigma = 4.07$ kHz around $f = 2.749690$ during feedback.

\indent To monitor the complete magnetic field vector $\vec{B}$ = $(B_x,B_y,B_z)$ during experiments with the stabilization protocol interleaved, we measure $f_{+1\mathrm{A}},f_{+1\mathrm{B}},f_{+1\mathrm{C}}$ and $f_{+1\mathrm{D}}$ (see Fig. 1b main text). In a similar approach as in section \ref{sec:SOM-fitting}, we fit a parabola to the four frequencies and use the residuals consisting of the difference of the four fitted frequencies and the corresponding frequencies obtained from equation \eqref{eq:S_Hp1} to perform a least-squares minimization to find $\vec{B}$.

The obtained values for $B_x$, $B_y$ and $B_z$ are shown in Figs. \ref{FigureS7}b,c,d. The standard deviations of these histogram verify a magnetic field stability during this protocol with $\sigma_{B_x}$ = 20 mG, $\sigma_{B_y}$ = 16 mG and $\sigma_{B_z}$ = 3 mG. Note that the average values of $\vec{B}$ from Supplementary Fig. \ref{FigureS7} show a discrepancy with the values obtained from the fit in the main text and section \ref{sec:SOM-fitting}, indicating that these measurements have been taken under slightly different conditions.

\FloatBarrier

\section{SUPPLEMENTARY NOTE: Coherence times of S3/S4 in $\ket{+1,\mathrm{D}}$}
\label{sec:S3S4_Coherence}

We use the same experimental sequence as in Fig. 5 of the main text to measure the $T_1$, $T_2^*$ and $T_2$ of the electron and nitrogen spin of S3/S4 (see Supplementary Fig. \ref{FigureS1}a). The observed coherence times are presented in Supplementary Table \ref{tab:S3S4_Coherences}, where also the coherence times of S1/S2 as measured in the main text are shown for completeness.

\begin{center}
\begin{figure*}[h]
\includegraphics{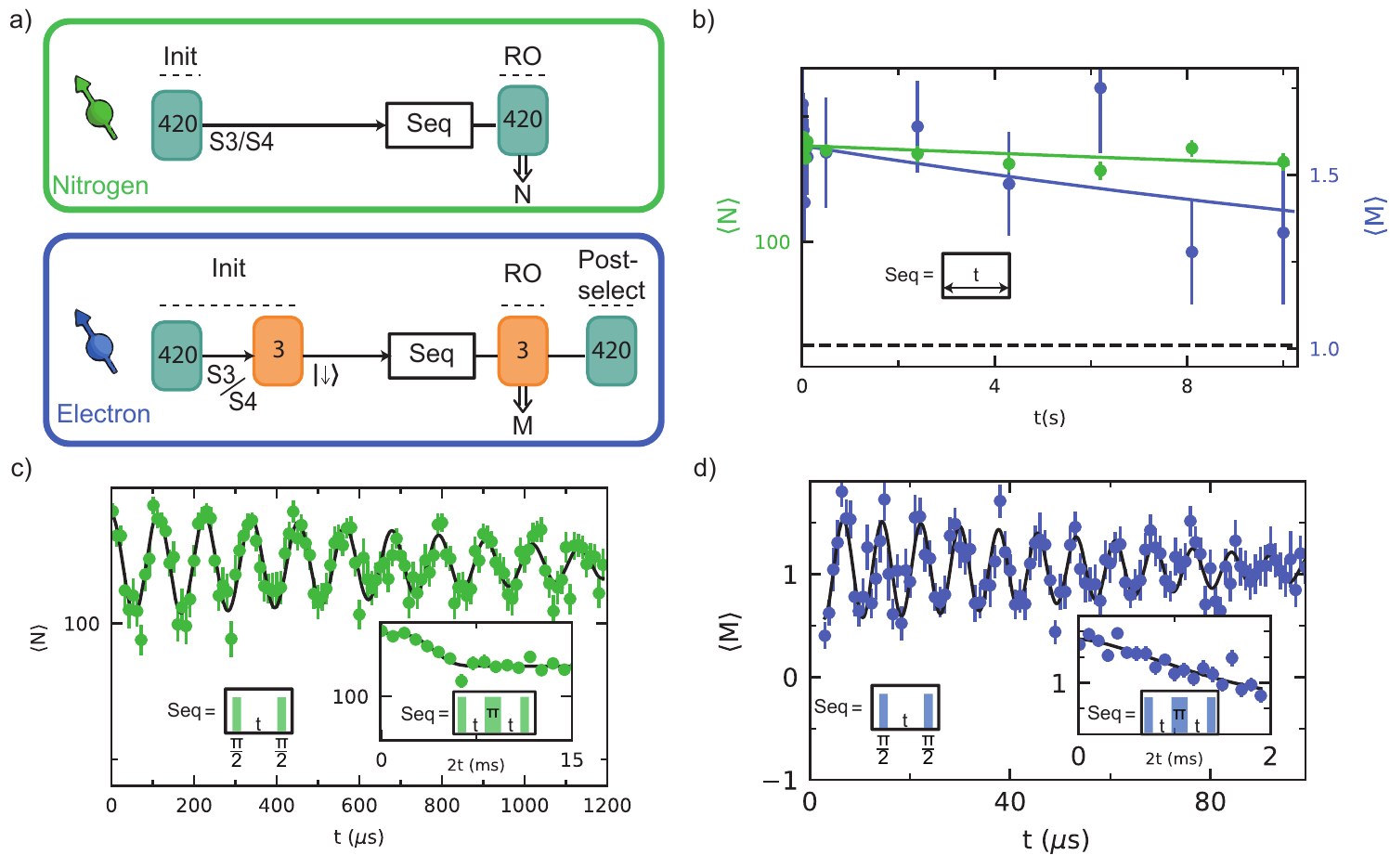}
\caption{\label{FigureS1}
\textbf{Timescales S3/S4.} (a) Sequence for initialisation of S3/S4 in $\ket{+1,\mathrm{D}}$ (top) and sequence for initialisation of all degrees of freedom of S3/S4 (bottom). These sequences are used in b,c and d. b) Relaxation of a combination of: the nitrogen state, JT axis and charge state (green), and only the electron spin state (blue). We fit (solid lines) both curves to $o + A_0 e^{-t/T}$. c) Ramsey and spin echo (inset) experiments on the nitrogen spin. d) Ramsey and spin echo (inset) experiments on the electron spin. The obtained coherence times are shown in Supplementary Table \ref{tab:S3S4_Coherences}. See Data analysis (Methods) for the complete fit functions of the Ramsey (equation 11) and spin echo (12) experiments presented in this figure.}
\end{figure*}
\end{center}
\FloatBarrier

\begin{table}[h!]
  \centering
  \begin{tabular}{|c|c|c|}
     \hline
                    & S3/S4 & S1/S2 \\ \hline
    $T_{\ket{+1,D}}$ & 104(38) s & 40(4) s\\ \hline
    $T_{1e}$ & 26(20) s & 21(7) s\\ \hline
    $T^*_{2e}$ & 82(10) $\upmu$s & 50(3) $\upmu$s \\ \hline
    $T^*_{2N}$ & 1.06(9) ms & 0.201(9) ms \\ \hline
    $T_{2e}$ & 1.5(1.2) ms & 1.00(4) ms \\ \hline
    $T_{2N}$ & 4.5(4) ms & 4.2(2) ms \\ \hline
   \end{tabular}
   \caption{\label{tab:S3S4_Coherences} Measured coherence and relaxation times of S3/S4 and S1/S2. For experimental data, see Supplementary Fig. \ref{FigureS1} and main text Fig. 5 respectively.}
\end{table}
\FloatBarrier

\section{SUPPLEMENTARY NOTE: Effective gyromagnetic ratio and spin coherence}
\label{sec:eff_gfactor}
\begin{center}
\begin{figure*}[h]
\includegraphics{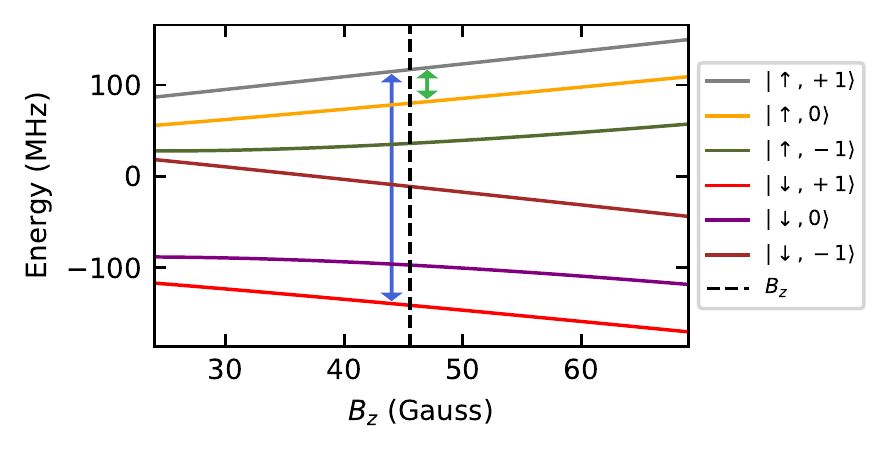}
\caption{\label{FigureS9}
\textbf{Energy levels for a single P1 center in JT axis D}. Simulation of the six energy levels which are labelled according to their P1 electron and nitrogen spin state. The black dashed line denotes the experimental magnetic field value. The green (blue) double headed arrow indicates the transition used in the nitrogen (electron) coherence experiments (Fig. 5c,d)}
\end{figure*}
\end{center}

\noindent Here we consider the effect of the electron-nuclear spin mixing due to a relatively large perpendicular hyperfine component ($\gamma_e |\vec{B}| \sim A_{\bot}$) on the expected coherence times. First we calculate the effective gyromagnetic ratios (labelled $\gamma_{n_{\mathrm{eff}}}$ and $\gamma_{e_{\mathrm{eff}}}$) of the two transitions used in the experiments of Fig. 5c, d (main text). We consider the six energy levels of a single P1 center distorted along the JT axis D at the experimental magnetic field, see dashed line Supplementary Fig. \ref{FigureS9}. Subsequently, we investigate the susceptibility of the energy levels to each component ($B_i$) of the magnetic field vector ($\vec{B}$). We then determine the tangent of both energy levels connected to the green (blue) double headed arrows at the experimental value of $B_i$ (see Supplementary Fig. \ref{FigureS9}) and calculate $\gamma_{n_{\mathrm{eff}},i}$ ($\gamma_{e_{\mathrm{eff}},i}$) as the difference between two tangents. The green (blue) arrow indicates the transition used in Fig. 5c(d) in the main text. We find $\gamma_{n_{\mathrm{eff}},z}$ ($\gamma_{e_{\mathrm{eff}},z}$) is more than 10 times larger than $\gamma_{n_{\mathrm{eff}},x/y}$ ($\gamma_{e_{\mathrm{eff}},x/y}$), and therefore we will only consider $\gamma_{n_{\mathrm{eff}},z}$ and $\gamma_{e_{\mathrm{eff}},z}$.

With the approach above, we obtain $\gamma_{n_{\mathrm{eff}},z}$ = $2\pi \cdot$ 0.206 MHz/G ($\sim$700$\gamma_n$) and $\gamma_{e_{\mathrm{eff}},z}$ = $2\pi \cdot$ 2.60 MHz/G ($\sim$0.93$\gamma_e$). Defining $\Gamma = \gamma_{e_{\mathrm{eff}},z}$/$\gamma_{n_{\mathrm{eff}},z}$, we find $\Gamma$ = 12.6. Thus, based on the spin mixing, we would expect the nitrogen spin coherence to be a factor 12.6 larger than the electron spin coherence. For S3/S4, we find a ratio $T^*_{2N}$/$T^*_{2e}$ $\sim$ 12 between the nitrogen and electron coherence (see Supplementary Table \ref{tab:S3S4_Coherences}), which agrees well. However, for spin S1 and S2 we find $T^*_{2N}$/$T^*_{2e}$ $\sim$ 4. The remaining discrepancy by a factor 3 is not yet understood.

\section{SUPPLEMENTARY NOTE: Entanglement sequence}
\label{sec:entanglement}
\noindent In this section the entanglement generation sequence for S1 and S2 is explained in detail. We begin by initializing S1 and S2 in $\ket{\Psi}_{init}$ = $\ket{\uparrow,+1,\mathrm{D}}_{S1}$ $\ket{\downarrow,+1,\mathrm{A}}_{S2}$ = $\ket{\uparrow \downarrow}$. Next we apply $U_1$ = $R_x(\pi/2)_{S1}R_x(\pi/2)_{S2}$, consisting of two $\frac{\pi}{2}$ pulses along $x$ to obtain the state:
\begin{equation}
    \label{eq:state1}
    U_1\ket{\Psi}_{init} = \frac{1}{2}\big( - i \ket{\uparrow \uparrow} + \ket{\uparrow \downarrow} - \ket{\downarrow \uparrow} - i \ket{\downarrow \downarrow} \big) 
\end{equation}

\noindent This state evolves for time $t$ and accumulates phase due to dipolar coupling $J$, as $U_{zz}(t)$ = $e^{-i 2 \pi \cdot \frac{J}{h} S_zS_z t}$:

\begin{equation}
    U_{zz}U_1 \ket{\Psi}_{init} = \frac{1}{2} e^{i \pi \cdot \frac{J}{h}\frac{t}{2} } \big( -i \cdot e^{-i \pi \cdot \frac{J}{h} t} \ket{\uparrow \uparrow} + \ket{\uparrow \downarrow} - \ket{\downarrow \uparrow} - i \cdot e^{-i \pi \cdot \frac{J}{h} t} \ket{\downarrow \downarrow} \big)
\end{equation}

\noindent Thereafter, we apply $U_2$ = $R_x(\pi)_{S1}R_x(\pi)_{S2}$, consisting of two $\pi$ pulses along $X$, followed by another free evolution time $t$:

\begin{equation}
    \ket{\Psi}_{final} = U_{zz}U_2U_{zz}U_1 \ket{\Psi} =  \frac{1}{2} e^{i \pi \cdot \frac{J}{h} t} \big( i \cdot e^{-i \pi \cdot \frac{J}{h}2t}\ket{\uparrow \uparrow} + \ket{\uparrow \downarrow} - \ket{\downarrow \uparrow} + i \cdot e^{-i \pi \cdot \frac{J}{h} 2t} \ket{\downarrow \downarrow} \big)
\end{equation}

\noindent Rewriting $\ket{S2}$ in the $X$ basis, where $\ket{+}$ = $\frac{\ket{\uparrow} + \ket{\downarrow}}{\sqrt{2}}$ and $\ket{-}$ = $\frac{\ket{\uparrow} - \ket{\downarrow}}{\sqrt{2}}$, results in:

\begin{equation}
\scalebox{1}{ $\ket{\Psi}_{final} = \frac{1}{2}e^{i \pi \cdot \frac{J}{h} t} \Bigg[ \bigg( \frac{i \cdot e^{- i \pi \frac{J}{h} 2t} + 1}{\sqrt{2}} \bigg) \ket{\uparrow +} + \bigg( \frac{i \cdot e^{-i \pi \cdot \frac{J}{h} 2t} -1 }{\sqrt{2}} \bigg) \ket{\uparrow -} + \bigg( \frac{i \cdot e^{- i \pi \cdot \frac{J}{h} 2t}-1}{\sqrt{2}} \bigg) \ket{\downarrow +} + \bigg(\frac{-i \cdot e^{-i \pi \cdot \frac{J}{h} 2t} -1}{\sqrt{2}} \bigg) \ket{\downarrow -} \Bigg]$ }
\end{equation}

\noindent Note that at time $2t$ = $\frac{1}{2|\frac{J}{h}|}$, and a negative coupling $J$, this yields to the entangled state: 

\begin{equation}
    \ket{\Psi}_{final} = -e^{\frac{i \pi}{4}} \frac{ \ket{\uparrow -} + \ket{\downarrow +}}{\sqrt{2}}
\end{equation}

\section{SUPPLEMENTARY NOTE: Optimization of initialisation/readout}
\label{sec:SOM-P1_INIT_opt}

\subsection{$^{14}$N and JT state}

\noindent Here we explain the optimization of single-shot readout and initialisation of the $^{14}$N and JT state of individual P1 centers. We define the fidelity of initialisation of S1 in $\ket{+1,\mathrm{D}}$ and S2, S3/S4 not in that state as

\begin{equation}
    F_{S1} = P(N(k+1)>N_{RO}|N(k)>N_{S1}),
\end{equation}

\noindent whereas for a mixture of all other possibilities we define

\begin{equation}
     F_{notS1} =  P(N(k+1)\leq N_{RO}|N(k)\leq N_{notS1}).
\end{equation}

\noindent In both cases $P(X|Y)$ is the probability to obtain X given Y. We define the combined initialisation and readout fidelity of these two cases as:

\begin{equation}
\label{eq:F_S1notS1_SOM}
    F = \frac{F_{S1} + F_{notS1}}{2}.
\end{equation}

\noindent The correlation plot between two subsequent measurements of $K$ = 820 binned DEER sequences is shown in Supplementary Fig. \ref{SFig:initreadout}a. First, we find a threshold that separates S1 from other P1 centers by setting $N_{S1}$  = $N_{notS1}$ = $N_{RO}$ all equal to $N_{sweep}$ and calculate the combined initialisation and readout fidelity $F$ while we sweep this parameter. A local maximum $N_{opt}$ = 477 is found separating spins S1 and S2 as shown in Supplementary Fig. \ref{SFig:initreadout}b (red dashed line). Second, we set $N_{RO}$ = $N_{notS1}$ = $N_{opt}$ to distinguish between \textit{S1} and \textit{not S1} and vary $N_{S1}$ as shown in Supplementary Fig. \ref{SFig:initreadout}c. As a trade-off between the success rate and fidelity $F$ we choose $N_{S1}$ = 522 to maintain a success probability $>$ 0.2. For Fig. 2f in the main text we have thus used $N_{notS1}$ = $N_{RO}$ = 477 and $N_{S1}$ = 522.

\begin{figure*}[h]
\includegraphics[scale=0.9]{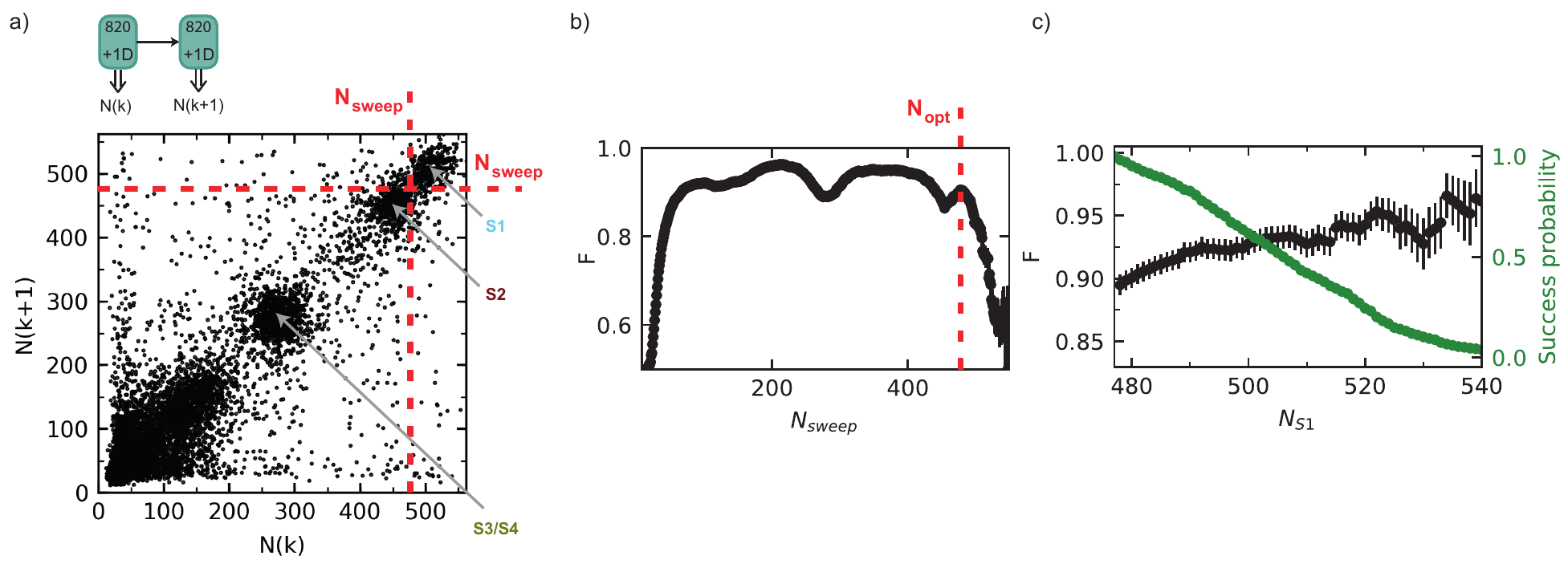}
\caption{\label{SFig:initreadout}
\textbf{Optimization of combined initialisation and readout of S1 in $\boldsymbol{\ket{+1,\mathrm{D}}}$}. \textbf{a)} Full correlation plot for consecutive measurement outcomes $N(k)$ and $N(k+1)$, both for $\ket{+1,\mathrm{D}}$ (same dataset as Fig. 2e in the main text). We set the thresholds $N_{S1}$  = $N_{notS1}$ = $N_{RO}$ all equal to $N_{sweep}$ (see main text Fig. 2e). Using these thresholds we calculate $F$ as in eq. \ref{eq:F_S1notS1_SOM} for different values of $N_{sweep}$. \textbf{b)} Fidelity $F$ as a function of $N_{sweep}$. The red dashed line indicates a local maximum of $F$ (here $N_{sweep}$ = $N_{opt}$ = 477) that optimally separates between S1 and S2 under the given constraints ($N_{S1}$  = $N_{notS1}$ = $N_{RO}$ = $N_{sweep}$). The same value ($N_{sweep}$ = $N_{opt}$) is shown by the red dashed lines in (a). \textbf{c)} Further improvement of $F$ at the cost of experimental rate is achieved by a stricter initialisation threshold $N_{S1}$. First we set $N_{notS1}$ = $N_{RO}$ = $N_{opt}$ and vary $N_{S1}$. The fidelity $F$ is plotted as function of $N_{S1}$ (black). The success probability of initialisation of $S1$ in $\ket{+1,\mathrm{D}}$ as compared to when $N_{S1}$ = 477 is plotted in green as a function of $N_{S1}$.}
\end{figure*}
\FloatBarrier

\subsection{Electron spin initialisation and readout}

\label{subsec:electron_init}

We initialize and measure the electron spin state of P1 centers through a DEER(y) sequence following initialisation of the $\ket{+1,\mathrm{D}}$ or $\ket{+1,\mathrm{A}}$ state. Again, we use the correlation of consecutive measurements $M(k)$ and $M(k+1)$ to determine the combined initialisation and readout fidelity. First, we define the fidelity of preparing $\ket{\uparrow}$ as

\begin{equation}
    F_{\ket{\uparrow}} = P(M(k+1)>M_{RO}|M(k)>M_{\ket{\uparrow}}),
\end{equation}

\noindent and the fidelity of preparing $\ket{\downarrow}$ as

\begin{equation}
     F_{\ket{\downarrow}} =  P(M(k+1)\leq M_{RO}|M(k)\leq M_{\ket{\downarrow}}).
\end{equation}

\noindent Again, the combined initialisation and readout fidelity is given as

\begin{equation}
    F_{\ket{\uparrow}/\ket{\downarrow}} = \frac{F_{\ket{\uparrow}} + F_{\ket{\downarrow}}}{2}.
\end{equation}

\begin{figure*}[h]
\includegraphics{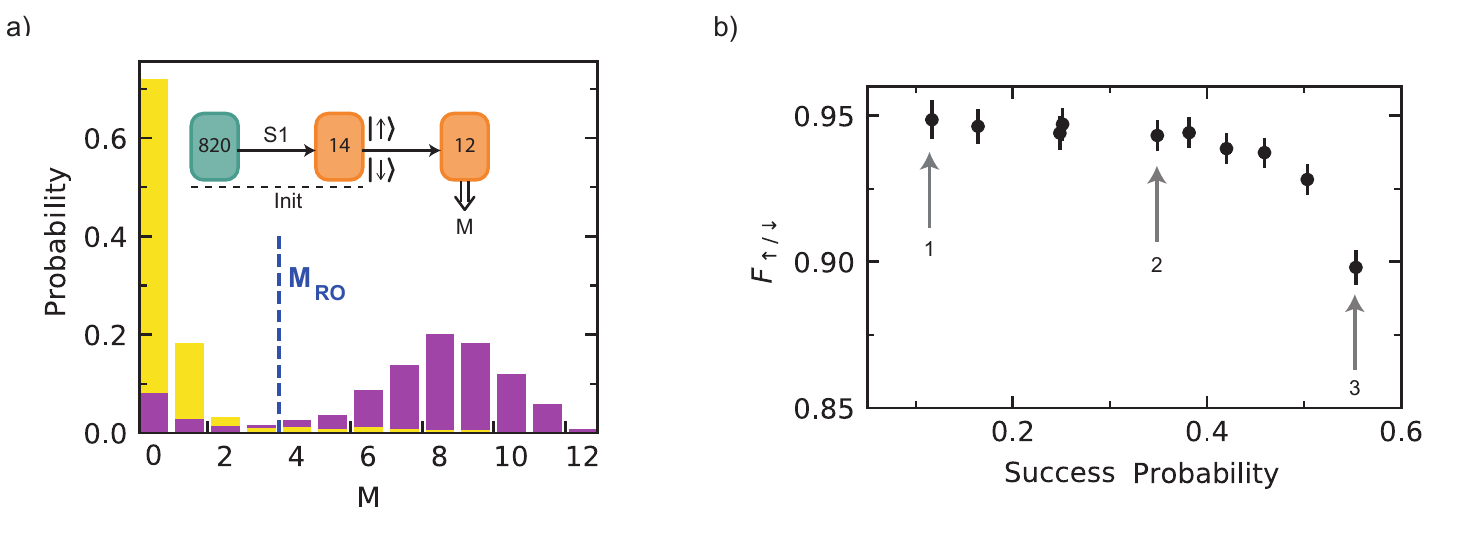}
\caption{\label{SFig:initreadout_spin}
\textbf{Optimization of combined electron initialisation and readout fidelity.} We trade off fidelity and experimental success rate by setting the thresholds and bin sizes of the DEER(y) sequences used. For optimization, a large range of thresholds and bin sizes are explored.  \textbf{a)} An exemplary case of probability distributions after initialisation. Here, after preparation of S1 in $\ket{+1,\mathrm{D}}$, the electron spin is initialized through a DEER(y) ($L$ = 14) with thresholds $M_{\ket{\uparrow}}$ ($>1$) and $M_{\ket{\downarrow}}$ ($\leq$1) before reading out ($L$ = 12). The parameters in this exemplary case ensure a larger success probability ($\approx$ 0.55) but lower $F_{\ket{\uparrow}/\ket{\downarrow}}$ = 0.90(1) as compared to those in Fig. 3d of the main text. \textbf{b)} The fidelity $F_{\ket{\uparrow}/\ket{\downarrow}}$ against the success probability of initialisation in $\ket{\uparrow}$. We calculate $F_{\ket{\uparrow}/\ket{\downarrow}}$ for a range of initialisation and readout bin sizes (L) and thresholds $M_{\ket{\downarrow}}$, $M_{\ket{\uparrow}}$ and $M_{RO}$. Only the maximum fidelities are shown, grouped over 10 intervals between 0.1 and 0.6 success probability. To ensure enough statistics, we only include intervals above 0.1 success probability. Numbered arrows indicate example cases: (1) parameters and probability histogram as shown in Fig. 3d in the main text; (2) $L$ = 14 for initialisation, $L$ = 8 for readout, $M_{\ket{\uparrow}}$ ($>8$), $M_{\ket{\downarrow}}$ ($\leq$1) and $M_{RO}$ (= 2); (3) corresponds to panel (a).}
\end{figure*}

\FloatBarrier

\noindent We calculate $F_{\ket{\uparrow}/\ket{\downarrow}}$ for a large range of initialisation and readout bin sizes (L) and thresholds ($M_{\ket{\downarrow}}$, $M_{\ket{\uparrow}}$ and $M_{RO}$). In Supplementary Fig. \ref{SFig:initreadout_spin}b maximum values of $F_{\ket{\uparrow}/\ket{\downarrow}}$ are shown as a function of success probability. We trade-off $F_{\ket{\uparrow}/\ket{\downarrow}}$ against success probability to obtain a combined initialisation and readout fidelity of $F_{\ket{\uparrow}/\ket{\downarrow}}$ = 0.95(1) while maintaining a success probability above 0.1.

\subsection{Optimization of sequential initialisation in $\ket{+1, \mathrm{D}}$ and $\ket{+1, \mathrm{A}}$}

\noindent For entanglement generations of S1 and S2 we optimize for high total initialisation and readout fidelity while maintaining a fast experimental rate. To increase the experimental rate, we use fast DEER measurement ($K$ = 5) on both $\ket{+1, \mathrm{D}}$ and $\ket{+1, \mathrm{A}}$ to quickly determine the likelihood of S1 and S2 in these states and only continue if the likelihood is high. Once this has been passed, we first continue to initialize S1 with a total of $K$ = 820 DEER sequences in $\ket{+1,\mathrm{D}}$. As S1 is now initialized, and S2 is more easily distinguished from S3/S4 it requires only $K$ = 50 DEER sequences to initialize S2 in $\ket{+1,\mathrm{A}}$. As consecutive measurements can disturb earlier prepared states, the small number of repetitions for the initialisation of S2 is beneficial for the overall fidelity. 

For electron spin initialisation and readout we optimize in a similar way as done in Subsection \ref{subsec:electron_init}. However, we now introduce additional measurements in order to take decoherence during sequential measurements into account and find a best combined initialisation and readout fidelity for both spins. Finally, we optimize the sequential initialisation (S1 in $\ket{\uparrow}$ then S2 in $\ket{\downarrow}$) for a high rate and fidelity but minimal disturbance of the initialized state of the other P1 center.

\section{SUPPLEMENTARY NOTE: NV fluorescence rate reference}

\noindent In this section, we verify that the discrete jumps in the DEER time traces (Fig. 2a of the main text) are not due to the changes in the detected fluorescence rate of the NV itself (e.g. due to ionization, spatial or spectral drifts). As shown in Supplementary Fig. \ref{fig:ionization_ref}, we repeatedly apply the DEER sequence while alternating between opposite readout bases. We combine these measurements to obtain a continuous DEER signal and a reference of the detected NV fluorescence rate. The DEER signal shows discrete jumps. These jumps are not observed in the reference signal, thus excluding that they are caused by changes in the detected NV fluorescence. 

\begin{figure*}[h]
\includegraphics[scale=1]{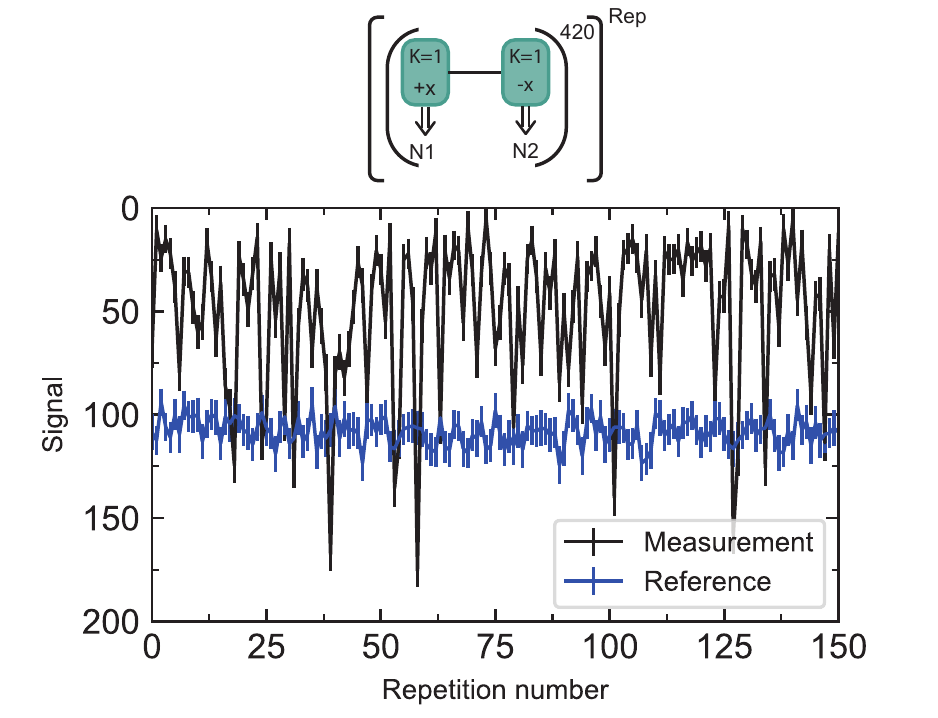}
\caption{\label{fig:ionization_ref} \textbf{Continuous DEER signal with reference signal.} Top: Experiment in which we alternate between a single DEER sequence ($K$=1, $f_{+1\mathrm{D}}$ as in main text Fig. 2b) with phase $+x$ and one with $-x$ for the final $\pi$/2 pulse. The outcomes $N_1$ and $N_2$ are summed in a bin size of 420 to obtain $N_{bin1}$ and $N_{bin2}$. The reference signal for the NV fluorescence detection rate is defined as $N_{ref}$ = ($N_{bin1}$+$N_{bin2}$)/2. The measurement signal is given as (2$\langle N_{ref} \rangle$ - $N_{bin1}$ + $N_{bin2}$)/2, here $\langle N_{ref} \rangle$ is the mean over the full dataset. Bottom: the measurement signal shows similar discrete jumps as in Fig. 2a of the main text, while the reference measurement remains approximately constant over time.}
\end{figure*}

\FloatBarrier

\bibliographystyle{naturemag}
\bibliography{SOM}